\documentclass[useAMS,usenatbib,twocolumn]{mnras}

\usepackage{amsmath,amssymb,mathrsfs,graphicx,float,latexsym,url}
\usepackage{epstopdf,ragged2e}
\usepackage{natbib,url}
\usepackage{hyperref}
\usepackage{widetext}

\usepackage[usenames,dvipsnames]{color}
\definecolor{darkred}{rgb}{0.5,0,0}
\definecolor{darkgreen}{rgb}{0,0.5,0}
\definecolor{darkblue}{rgb}{0,0,0.5}
\definecolor{prussian}{rgb}{0.0, 0.19, 0.33}
\definecolor{richelectricblue}{rgb}{0.03, 0.57, 0.82}
\definecolor{teal}{rgb}{0.0, 0.5, 0.5}
\definecolor{mediumseagreen}{rgb}{0.24, 0.7, 0.44}
\definecolor{lust}{rgb}{0.9, 0.13, 0.13}
\definecolor{ballblue}{rgb}{0.13, 0.67, 0.8}
\definecolor{darkcyan}{rgb}{0.0, 0.55, 0.55}
\definecolor{mountainmeadow}{rgb}{0.19, 0.73, 0.56}
\definecolor{palecarmine}{rgb}{0.69, 0.25, 0.21}
\definecolor{richcarmine}{rgb}{0.84, 0.0, 0.25}
\definecolor{tangelo}{rgb}{0.98, 0.3, 0.0}
\definecolor{venetian}{rgb}{0.784,0.031,0.082}
\definecolor{bdfrance}{rgb}{0.192,0.549,0.906}
\usepackage{mathtools}
\usepackage{amsbsy}
\usepackage{bm}
\usepackage{float}

\newcommand{\aPM}{\alpha_{\rm PM}}

\newcommand{\bv}{\boldsymbol{v}}
\newcommand{\bb}{\boldsymbol{B}}
\newcommand{\bE}{\boldsymbol{E}}
\newcommand{\bj}{\boldsymbol{J}}

\newcommand{\Ms}{M_{\star}}
\newcommand{\Rs}{R_{\star}}

\def\apj{{ApJ}}

\def\apjl{{ApJL}}

\def\aap{{A\&A}}

\def\mnras{{MNRAS}}

\def\nat{{Nature}}

\def\prd{{Physical Review D}}
\def\prl{{Phys. Rev. Lett.}}

\def\04a{{2004 a}}
\def\04b{{2004 b}}

\begin{document}

\title[Magnetars in non-linear electrodynamics]
{Magnetar structure in non-linear electrodynamics with mixed poloidal-toroidal fields}
\author[Arthur G. Suvorov \& Jos{\'e} A. Pons]{Arthur G. Suvorov$^{1}$\thanks{arthur.suvorov@ua.es} and Jos{\'e} A. Pons$^{1}$ \\
$^1$Departament de F{\'i}sica Aplicada, Universitat d'Alacant, Ap. Correus 99, E-03080 Alacant, Spain\\
}

\date{Accepted ?. Received ?; in original form ?}

\pagerange{\pageref{firstpage}--\pageref{lastpage}} \pubyear{?}

\maketitle
\label{firstpage}

\begin{abstract}
\noindent Magnetars have inferred polar field strengths in excess of the Schwinger limit, where non-linear electromagnetic effects can be significant. Their internal fields may be even stronger, suggesting that Maxwellian characterizations of hydromagnetic structure may require revision. A generalized Grad-Shafranov equation, describing static and axisymmetric fluid stars with mixed poloidal-toroidal fields, is introduced and subsequently solved in a perturbative scheme to calculate quadrupolar deformations. In the Born-Infeld theory, we show that the toroidal field has a maximum strength set by the scale parameter, $b$, implying an upper limit to the stellar prolateness, $|\epsilon_{\rm max}| \sim 10^{-5} \left(b/10^{16}\text{ G}\right)^2$, that is independent of field specifics. Observations of magnetar phenomena that are interpreted as evidence for ellipticity, such as precession, can thus implicitly constrain post-Maxwellian parameters in a way that complements terrestrial experiments. Toroidal ceilings also have implications for dynamo theory and gravitational waves, which we revisit together with field evolution in crusts abiding by beyond-Maxwell physics.
\end{abstract}

\begin{keywords}
magnetohydrodynamics, stars: magnetars, magnetic fields, gravitational waves
\end{keywords}

\section{Introduction} \label{sec:intro}

The magnetar class of neutron star possess extremely strong magnetic fields. Their near-surface field strengths inferred from spindown \citep{mcg14}, cyclotron line \citep{tiengo13}, and X-ray spectral \citep{guv08} features categorically exceed the \cite{sch51} value ($B_{\rm QED} \approx 4.41 \times 10^{13}$~G), where quantum-electrodynamic (QED) processes acquire non-negligible cross sections \cite[see][for a review]{lai01}. Evidence for such processes, most notably vacuum birefringence, has been recently identified in polarimetric measurements from pulsed components in the magnetar 4U 0142+61 \citep{tav22} and other highly-magnetised sources, such as RX J1856.5--3754 \citep{mig17,tav24}. Magnetothermal \cite[e.g.,][]{deh23} and hydromagnetic \cite[e.g.,][]{cr13} modelling further suggests that core and/or crustal fields may locally exceed surface values by factors of at least a few, especially if there is a dominant toroidal component hidden within as inferred from X-ray \citep{mak14} and radio \citep{des24} observations. Understanding the impact of nonlinear electrodynamics (NLED) on magnetar structure, both internal and magnetospheric, may thus be important to accurately interpret multiwavelength data.

One typically expands the nonlinear Lagrangian out to post-Maxwellian (PM) or post-post-Maxwellian (PPM) order, in direct analogy with post-Newtonian expansions, to simplify calculations as the equations become more mathematically involved due to added nonlinearities \cite[e.g.,][]{den04}. Analyses of magnetar phenomena accounting for NLED effects have been performed using these expansions, including that related to: augmentations of the Poynting flux (and hence spindown) due to effective changes of the vacuum permeability \citep{den16,heyl19}, magnetospheric structure \citep{petri15,li20}, and gravitational-wave (GW) emission from stars with purely toroidal fields \citep{per18}. Thus far however, no studies explicitly tackling the NLED-adjusted equations of hydromagnetic equilibrium have been carried out at either PM or exact levels. The goal of this work is to fill this gap. We construct and perturbatively solve a `generalised Grad-Shafranov' equation applying to stationary and axisymmetric plasmas, where the toroidal field can be implicitly expressed in terms of an arbitrary function of the poloidal flux.

Allowing for mixed fields reveals that the toroidal field is quantitatively capped by the PM parameter(s) because the current density is bounded. The toroidal ceiling persists when using the full nonlinear Lagrangian, at least in the case of the \cite{bi34} (BI) theory which forms our primary focus. This result, which seems to have gone unnoticed thus far, has the immediate implication that toroidally-dominated configurations cannot exist in the interior of magnetars unless the BI scale parameter, $b$, is sufficiently large. Using the non-barotropic approximation introduced by \cite{mast11} and \cite{akg13}, we calculate the deformation induced by mixed fields in the interior and show there is a maximum prolateness set by the scale parameter. We thus argue that the GW detectability of deformed magnetars is more pessimistic in NLED. 

Another consequence of toroidal capping is that models of magnetar phenomena stipulating strong toroidal fields can be used to implicitly constrain NLED parameters, at least if the surface field is known. For example, fits to gamma-ray burst (GRB) afterglows \cite[e.g.,][]{lasky16,suvk20} or long period X-ray modulations \cite[e.g.,][]{mak14} often favour toroidally-dominated configurations. These constraints can complement those coming from terrestrial experiments, such as the Polarizzazione del Vuoto con LASer \cite[PVLAS;][]{ej20} and Large Hadron Collider \cite[LHC;][]{ellis17}. We argue, for instance, that a prolateness of $\sim 10^{-3}$ mandates that the BI scale parameter $b \gtrsim 10^{17}$~G; a factor $\sim 3$ tighter than the bound obtained by \cite{akman18} using hydrogen ionisation data.

We take the opportunity to also revisit magnetar evolution in NLED more generally, such as regards Hall-Ohm diffusion in the crust due to a generalised Amp{\'e}re's law. It is shown that because the electric current is effectively reduced in a manner proportional to the PM parameter(s), ultra-strong crustal fields may decay at a \emph{slower} rate than the Maxwell case, with implications for cooling and field longevity. Adjustments to the magnetic stress-energy tensor may also influence the rate and geometry of failure events in the crust \citep{perna11,lg19,koj24}.

This work is organised as follows. We review the ingredients of magnetohydrodynamics (MHD) including a general electromagnetic Lagrangian in Section~\ref{sec:mhd}, with equations for static and axisymmetric interiors written in a `Grad-Shafranov' form (Sec.~\ref{sec:stataxi}), focussing on the BI (Sec.~\ref{sec:bilag}) and general PM (Sec.~\ref{sec:postmaxwell}) cases. Perturbative solutions for the vacuum exterior are then constructed (Sec.~\ref{sec:exterior}) as a step towards the interior problem (Sec.~\ref{sec:magfield}). The quadrupolar ellipticity is calculated in Sec.~\ref{sec:quadellip}, with implications for the GW luminosity of magnetars explored in Sec.~\ref{sec:gwdetect}. Various models for phenomena that invoke strong toroidal fields are then revisited in NLED in Section~\ref{sec:magphenom}, with discussion and avenues for extensions provided in Section~\ref{sec:discussion}. 

\section{Magnetohydrodynamics with nonlinear electrodynamics} \label{sec:mhd}

In the classical description of magnetised fluids, the electrodynamic sector is derived by varying the Maxwell density, $\mathcal{L}_{\rm M} = -\tfrac{1}{4} F_{\mu \nu} F^{\mu \nu} \equiv - F^2$, where $\boldsymbol{F}$ is the Faraday tensor\footnote{The reader is cautioned that some authors use $F$ or $F^2$ to denote $F_{\mu \nu} F^{\mu \nu}$ directly \cite[e.g.,][]{heyl19} while others, including us here, incorporate the factor 1/4. One can, in general, also include the current term in the action.}. In a general theory, the Lagrangian can depend on not only $F^2$ but also its electromagnetic dual, $-G^2 \equiv -\tfrac{1}{4} \left( \star F \right)_{\mu \nu} F^{\mu \nu}$, and one can work with $\mathcal{L}(F^2,G^2)$. However, we consider cases with Lagrangian functions independent of $G$, writing $L(F^2)$. This is because (i) $G$ appears at higher orders (relative to $F$) in BI theory (see Sec.~\ref{sec:bilag}), (ii) $G$ vanishes identically in the ideal MHD limit where the electric field, $\bE$, is zero, and more generally (iii) in keeping with (non-ideal) MHD assumptions we have $\bE \sim (v/c)\bb \ll \bb$ for bulk plasma velocity $\bv$, and hence $G^2 = \bE \cdot \bb \ll F^2 = (\bb^2- \bE^2) /2$, where $\bb$ is the magnetic field expressed in terms of orthonormal tetrad components.

In NLED MHD with a single fluid, the current density, $\bj$, is found through the generalised Amp{\'e}re law in the limit where high-frequency plasma excitations are ignored \cite[e.g.,][]{per18},
\begin{equation} \label{eq:ampere}
-\nabla \times \left[ L'(F^2) \bb \right] = \frac{4 \pi}{c} \bj ,
\end{equation}
where the prime denotes differentiation with respect to argument. The Lorentz force takes its usual form but with the modified current \eqref{eq:ampere} \cite[e.g.,][]{boil70,kim00,li20},
\begin{equation} \label{eq:lorentz}
\boldsymbol{f}_{\rm mag} \equiv \frac{1}{c} \boldsymbol{J} \times \boldsymbol{B} = -\frac{\nabla \times \left[ L'(F^2) \bb \right] \times \bb}{4\pi}.
\end{equation}
Using the generalised force \eqref{eq:lorentz}, the equations of motion for a magnetised fluid can be written at any desired level of complexity. 

\subsection{Static, axisymmetric equilibria} \label{sec:stataxi}

We restrict our attention to axisymmetric, static, and ideal fluids, so that the force-balance equations simplify considerably (the latter are relaxed in Sec.~\ref{sec:magphenom}). Given that NLED corrections are most likely to apply, if at all, to magnetar environments, rotational deformations are expected to be small except perhaps in the first year of life as the huge magnetic field spins down the star rapidly: the observed Galactic magnetars have spin periods on the order of several seconds \citep{mcg14}. This justifies the choice $\bv = 0$, at least for mature objects, even if it is strictly-speaking inconsistent to model mixed poloidal-toroidal magnetic fields in the absence of meridional flows \citep{col08,suvg23}.

With these restrictions and using expression \eqref{eq:lorentz}, the equations of motion read
\begin{equation} \label{eq:eom0}
\nabla p + \rho \nabla \Phi = \boldsymbol{f}_{\rm mag},
\end{equation}
for mass density $\rho$, pressure $p$, and gravitational potential $\Phi$, related via the Poisson equation,
\begin{equation} \label{eq:poisson}
\nabla^2 \Phi = 4 \pi G \rho,
\end{equation}
where $G$ is Newton's constant. The system is closed via an equation of state (EOS; see Sec.~\ref{sec:hydrostatics}). Other forces can be easily included within equation \eqref{eq:eom0}, such as those associated with crustal elasticity (see Sec.~\ref{sec:failures}). To proceed, we note that the solenoidal nature of $\bb$ allows us to write \cite[e.g.,][]{hask08,mast11,akg13}
\begin{equation} \label{eq:bfield}
\bb = \nabla \psi(r,\theta) \times \nabla \phi + T(r,\theta) \nabla \phi,
\end{equation}
where $\psi$ is the poloidal flux function, $T$ is a toroidal component, and we use spherical coordinates $\{r,\theta,\phi\}$. Expanding out \eqref{eq:eom0} we can write two equations, from the polar $(r,\theta)$ and azimuthal $(\phi)$ components, involving the functions $\psi$ and $T$ given some $L$. These equations take the remarkably simple form \cite[cf.][]{li20}
\begin{equation} \label{eq:poloidaleq}
\begin{aligned}
\nabla p + \rho \nabla \Phi =& \frac{\nabla\psi}{4 \pi r^2 \sin^2\theta} \left[ \tilde{\Delta} \psi + \frac{I(\psi) I'(\psi)}{L'(F^2)}  \right],
\end{aligned}
\end{equation}
and 
\begin{equation} \label{eq:toroidaleq}
I(\psi) = -L'(F^2) T(r,\theta),
\end{equation}
where we introduce a generalised Grad-Shafranov (GGS) operator\footnote{Note that this is actually the negative of the usual GS operator in the Maxwell limit where $L'(F^2) = -1$ \cite[e.g.,][]{mast11}.},
\begin{equation}
\tilde{\Delta} = \frac{\partial}{\partial r} \left[ L'(F^2) \frac{\partial}{\partial r} \right] + \frac{\sin\theta}{r^2} \frac{\partial}{\partial \theta} \left[ \frac{L'(F^2)}{\sin \theta} \frac{\partial}{\partial \theta} \right] .
\end{equation}
Equation \eqref{eq:toroidaleq} can be inverted to deduce $T$ for any given Lagrangian function $L$ using the fact that
\begin{equation} \label{eq:explicitf}
F^2 \equiv \frac{1}{2} \bb \cdot \bb =\frac{\nabla \psi \cdot \nabla \psi + T^2} {2 r^2 \sin^2 \theta}.
\end{equation} 
Clearly in the linear (Maxwell) limit, the toroidal function is identifiable exactly as a function of the poloidal flux, though in general it will be augmented by other terms which are complicated functions of $r$ and $\theta$. At least in the BI or post-Maxwellian theories, expression \eqref{eq:toroidaleq} can be inverted analytically (see Secs.~\ref{sec:bilag} and \ref{sec:postmaxwell}, respectively).

As in the Maxwell case, some (arbitrary) choice for $I$ must be made. A physically-motivated option in the spirit of the `twisted torus' models is set by demanding that the toroidal field be confined to the neutral curves of the poloidal field \citep{bs04,bs06}. This can be ensured by taking $I(\psi) \propto \left(|\psi| - \psi_{c}\right)^{\sigma}$ for $|\psi| > \psi_{c}$ and zero elsewhere, where $\psi_{c}$ and $\sigma$ are positive constants \cite[e.g.,][]{c09,urban23}. Physically, $\psi_{c}$ corresponds to the value of $|\psi|$ defining the last poloidal field line closing within the star. We take $\sigma = 2$ as in \cite{mast11} and thereby set
\begin{equation} \label{eq:toroidalchoice}
I(\psi) = i_{0} \left( |\psi| - \psi_{c} \right)^2 \Theta\left(|\psi| - \psi_{c}\right),
\end{equation}
where $\Theta$ is the Heaviside function and $i_{0}$ is some constant defining the relative strengths of the poloidal and toroidal sectors.

One is now tasked with solving the non-linear, elliptic equation \eqref{eq:poloidaleq} together with \eqref{eq:poisson} given some EOS and boundary conditions. In general, the gravitational potential must match to some multipole expansion at the surface, defined by the vanishing of the (hydromagnetic) pressure, $p=0$. The magnetic field should therefore match to some force-free exterior where the right-hand side of \eqref{eq:poloidaleq} vanishes. Unfortunately, in the general NLED case, the electrovacuum equations cannot be solved analytically even without currents ($i_{0}=0$). Deducing the boundary conditions therefore requires a numerical or perturbative approach. The latter construction, together with (approximate) solutions to equation~\eqref{eq:poloidaleq} more generally, are deferred to Sec.~\ref{sec:fieldstructure}. 

\subsection{Born-Infeld theory} \label{sec:bilag}

Thus far we have treated the function $L$ as general. In practice, one must select a theory in order to make numerical estimates. A number of theories have been motivated in the literature from QED or otherwise \cite[see][and references therein]{modmax20}, though we concentrate primarily on BI theory and post-Maxwellian ones more generally.

The BI theory has the simple, closed form Lagrangian \citep{bi34}
\begin{equation} \label{eq:bilagrangian}
L_{\rm BI}(F^2) = -b^2 \sqrt{1 + \frac{2F^2}{b^2}} + b^2,
\end{equation}
for scale parameter $b$, introduced to temper the singular behaviour of the classical electron self-energy. \cite{bi34} anticipated $b \approx 0.7 c^4 m_{e}^2/e^3 \approx 4 \times 10^{15}$~G, though larger values are required by modern experiments \cite[e.g.,][]{ej20}. In fact, we will argue that even $b \gtrsim 10^{16}$~G places severe restrictions on the hydromagnetic structure of magnetar interiors (see Sec.~\ref{sec:magfield}). Based on this, there is the tantalising prospect of constraining the BI parameter by comparing theoretical expectations with observed values of the more extreme magnetar phenomena; this topic is covered in Secs.~\ref{sec:gwdetect} and \ref{sec:magphenom}. 

\begin{figure}
\includegraphics[width=0.487\textwidth]{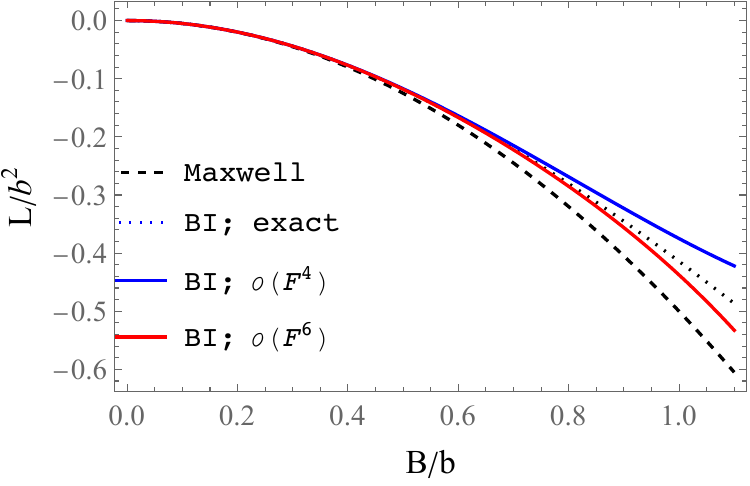}
\caption{Comparison between the (normalised) BI function \eqref{eq:bilagrangian}, at either exact (dotted, black), post-Maxwellian (solid, blue), or post-post-Maxwellian (solid, red) orders \eqref{eq:biexpansion}, as compared to the Maxwell case (dashed, black).
\label{fig:lei}}
\end{figure}

Figure~\ref{fig:lei} compares the Maxwell Lagrangian ($b \rightarrow \infty$) to the exact BI function \eqref{eq:bilagrangian} and its PM and PPM expansions, which can be read off from
\begin{equation} \label{eq:biexpansion}
L_{\rm BI}(F^2) =  -F^2 + \frac{F^4}{2 b^2} - \frac{F^6}{2 b^4} + \mathcal{O}(F^8) .
\end{equation}
We see that the PM and PPM expansions work remarkably well even up to $B \sim b$, and more importantly the latter at least provides a \emph{conservative} estimate for NLED impact around such values (compare red and dotted curves). 

The BI Lagrangian is sufficiently simple that expression \eqref{eq:toroidaleq} for the toroidal field can be inverted analytically, leading to
\begin{equation} \label{eq:toroidalbi}
T(r,\theta) = \frac{ I(\psi) \sqrt{b^2 r^4 \sin^2 \theta + r^2 \nabla \psi \cdot \nabla \psi}} {r \sqrt{b^2 r^2 \sin^2 \theta - I(\psi)^2}}.
\end{equation}
The formula \eqref{eq:toroidalbi} has important implications for the magnetic fields pertaining to magnetars, as it imposes the bound
\begin{equation} \label{eq:upperbound}
I(\psi)^2 < b^2 r^2 \sin^2\theta,
\end{equation}
else the toroidal field diverges and subsequently becomes \emph{imaginary}. Requiring that the toroidal flux function $I(\psi)$ not exceed some maximum value \eqref{eq:upperbound}, and also tend to zero around the poles and origin, implies a range of viable energy partitions (i.e., limits on $|i_{0}|$) for a given poloidal geometry, set by $\psi$; see Sec.~\ref{sec:magfield}. One may physically interpret inequality \eqref{eq:upperbound} by noticing that regularising the classical electron self-energy must set a maximum current, and hence maximum toroidal strength. In the toroidally-dominated limit, the right-hand side of \eqref{eq:toroidaleq} behaves as $\sim T/\sqrt{1+T^2/ b^2 r^2 \sin^2 \theta} \leq b r \sin\theta$, which cannot match $I(\psi)$ beyond some cutoff. A similar conclusion should hold in non-axisymmetry using a general Helmholtz decomposition for $\bb$.

\subsection{Parameterised post-Maxwellian theory} \label{sec:postmaxwell}

Rather than working with some specific theory, one can introduce a set of free parameters so as to consider a theory-agnostic range of corrections in the form
\begin{equation} \label{eq:ppm}
L(F^2) = -F^2 + \aPM F^4 + \alpha_{\rm PPM} F^6 + \mathcal{O}(F^8),
\end{equation}
for some hierarchy of parameters $\aPM$, $\alpha_{\rm PPM}$, ..., with units of $\text{Gauss}^{-2}$, $\text{Gauss}^{-4}$, and so on. These coefficients can be read off from Maclaurin expansions for any given theory.

At leading order, one can also invert expression \eqref{eq:toroidaleq} analytically using the cubic root formula. A critical feature of the solution is that there is similarly a square root term in the denominator of $T$ as in the BI case \eqref{eq:toroidalbi}, which behaves as
\begin{equation} \label{eq:pmtoroidal}
\begin{aligned}
T_{\rm PM}(r,\theta) \propto& \Big\{ \Big[ \aPM r^{4} \csc^2\theta I(\psi)^2 \\
&- \frac{4}{27} \left( r^2 - \aPM \csc^2\theta \nabla \psi \cdot \nabla \psi \right)^{3} \Big]^{1/2} \\ 
&- r^2 \aPM^{1/2} \csc \theta I(\psi) \Big\}^{-1/3}.
\end{aligned}
\end{equation}
The above implies a maximum limit to the toroidal field strength for a different reason than for the BI case, namely that $L'(F^2)$ can tend to zero. Of course, higher-order counterterms or non-polynomial turnovers could neutralise this feature. 

\section{Hydromagnetic structure} \label{sec:fieldstructure}

We are now in a position to tackle one of the main goals of this work: to estimate the impact of structural changes to magnetar interiors due to strong magnetic fields in NLED. 

\subsection{Force-free poloidal fields} \label{sec:exterior}

We begin with a simplified version of the problem, where the left-hand side of \eqref{eq:poloidaleq} is ignored together with currents, viz.
\begin{equation} \label{eq:forcefree}
0 =  \tilde{\Delta} \psi .
\end{equation}
A necessary step towards approaching the fluid problem involves solving for a force-free configuration via equation \eqref{eq:forcefree}, as the latter defines matching conditions at the stellar surface. Although the field in the interior of the star may be considerably larger and full Lagrangians should be used, the surface field in Galactic magnetars is sufficiently small such that the PM or PPM terms within \eqref{eq:ppm} should reasonably encapsulate the dynamics (again unless the theory has non-polynomial turnovers). For the remainder of this paper, we express the spherical radius $r$ in units of the stellar radius $R_{\star}$ (so that $r$ is dimensionless). 

In general, a complication of cases with $\aPM \neq 0$ is that the usual decomposition into multipolar components cannot be carried out because the angular symmetry is broken. To make progress, we follow the method described by \cite{c09} and \cite{suvg23} which involves projecting onto a Legendre basis. Consider the expansion,
\begin{equation}
\psi(r ,\theta) = -\sin \theta \sum_{\ell=1}^{\ell_{\rm max}} f_{\ell}(r) \frac {d P_{\ell} (\cos \theta)}{d \theta},
\end{equation}
for radial functions $f_{\ell}(r)$ to be solved for. Making use of the orthogonality relations
\begin{equation} \label{eq:orthog}
\int^{\pi}_{0} d \theta  \frac {d P_{\ell}(\cos \theta)} {d \theta} \frac { d P_{\ell'}(\cos \theta)} {d \theta} \sin \theta  = \frac{2 \ell \left(\ell +1 \right)}{2 \ell + 1} \delta_{\ell \ell'},
\end{equation}
one may project equation \eqref{eq:forcefree} into harmonics by multiplying by Legendre polynomials and then integrating over $\theta$. In this way we reduce the problem to a (coupled) set of $\ell_{\rm max}$ ordinary differential equations for the functions $f_{\ell}(r)$.

While this could be carried out for some arbitrary $\ell_{\rm max}$, we assume that the dipole component is dominant. Equation \eqref{eq:forcefree} then decomposes into a single ordinary differential equation for $f_{1}(r)$. Regrettably, the resulting equation does not appear to admit any closed-form solutions owing to the high degree of nonlinearity. If, however, we further adopt the ansatz
\begin{equation} \label{eq:expansion}
f_{1}(r) = f_{1,0}(r) + \aPM f_{1,1}(r) + \alpha_{\rm PM}^2 f_{1,2}(r) + \ldots,
\end{equation}
and expand out equation \eqref{eq:forcefree} by reading off the coefficients of $\alpha_{\rm PM}^{k}$ for each $k$, a series solution \emph{can} be found. Note that the Maxwell contribution takes the well-known form $f_{1,0} = B_{p}/2r$ for polar field strength $B_{p}$ (twice that of the equatorial value). Expanding out terms to successive orders, we find the solution {(see Appendix~\ref{sec:appb} for a derivation)}
\begin{equation} \label{eq:exact}
\begin{aligned}
f_{1}(r) =& \frac{B_{p}}{2} \left( \frac{1}{r} + \frac{\aPM}{30} \frac{B_{p}^2}{r^7} + \frac{43 \aPM^2}{3000} \frac{B_{p}^4}{r^{13}} \right) \\
&+ \mathcal{O}(\aPM^{3},\alpha_{\rm PPM}, \ldots),
\end{aligned}
\end{equation}
the leading-order piece of which has been reported before \cite[e.g.,][]{heyl19}. It is possible to obtain the full PM series solution \eqref{eq:expansion}, with $f_{1,k}(r) \propto B_{p}^{2k+1} r^{-6k-1}$. For a general $\ell$-pole, we instead find the scaling $f_{\ell,k}(r) \propto r^{-6k-\ell}$. 

Expression \eqref{eq:exact} demonstrates, as expected, that PM effects die off rapidly with radius relative to Maxwellian ones. At the stellar surface, the dimensionless, relative magnitude of the terms can be simply read off as
\begin{equation} \label{eq:ratio1}
\frac{f_{1,1}(1)}{f_{0,1}(1)} = \frac{1}{30} B_{p}^2 \aPM,
\end{equation}
which is $\ll 1$ even for $B_{p}^2 \aPM \sim 1$. Note if \eqref{eq:ratio1} were of order unity this would signal the breakdown of our perturbative approach, either at the level of the radial expansion \eqref{eq:expansion} or the action \eqref{eq:ppm}. The former can at least be quantified easily by examining the ratio $f_{1,2}(1)/f_{1,1}(1) = \frac{43}{100} \aPM  B_{p}^2$. For BI theory, we report here the PPM solution {(Appendix~\ref{sec:appb})}
\begin{equation} \label{eq:bio6}
\begin{aligned}
\psi_{\rm BI,PPM}(r,\theta) =& \frac{B_{p}}{2} \Big(\frac{1}{r} + \frac{1}{60 r^7}\frac{B_{p}^2}{b^2} \\
&+ \frac{31}{84000 r^{13}} \frac{B_{p}^4}{b^4} - \frac{5479}{7056000 r^{19}} \frac{B_{p}^6}{b^6} \Big),
\end{aligned}
\end{equation}
which will be useful for constructing internal fields in Sec.~\ref{sec:magfield}. For the Galactic magnetar SGR 1806--20 with $B_{p} \approx 4 \times 10^{15}$~G \citep{mcg14}, we see from expressions \eqref{eq:ratio1} and \eqref{eq:bio6} that the size of relative corrections are small with $\aPM \sim 5 \times 10^{-33} \text{ G}^{-2}$ if $b \sim 10^{16}$~G. Strictly speaking, the PPM corrections are no stronger than the PM ones for $B_{p} \lesssim 0.7 b$. {While a fully-numerical, grid-based method to solve \eqref{eq:forcefree} would hold the advantage that no such projections are necessary, the perturbative approach used here is reasonable because all Galactic magnetars satisfy this condition (though cf. Sec.~\ref{sec:grbs}). For the interior, hydromagnetic problem -- discussed next -- the local field may exceed $b$ and this may no longer be the case. On the other hand, since the toroidal maximum \eqref{eq:upperbound} is a \emph{non-perturbative} result and the field structures we find using either $\mathcal{O}(b^{-4})$ or $\mathcal{O}(b^{-6})$ expansions are quantitatively similar (Sec.~\ref{sec:magfield}), we expect that any mismatches with `true' solutions are subleading with respect to other approximations made here (e.g., Newtonian gravity).} 


We close this section with some remarks on normalisation. Although we have and continue to use the symbol $B_{p}$ to refer to the polar field strength, it is not really so in NLED. This can be seen by substituting $r=1$ into expressions \eqref{eq:exact} or \eqref{eq:bio6}, revealing an increase to the polar cap strength. Although we could tune the respective coefficients between the Maxwell and NLED cases to match (for instance) the values $\psi(1,\pi/2)$, this would instead mean mismatching the asymptotics (i.e. the value of $\underset{r \to \infty}{\lim} r \psi$). Such an issue is familiar in studies of general-relativistic extensions of {Maxwell's equations} \citep{rez04,petri15}. We thus refer to $B_{p}$ as the polar strength in either case, to be interpreted as the polar dipole moment divided by the cube of the stellar radius, as measured by an observer at infinity.

\subsection{Internal fields} \label{sec:magfield}

In a self-consistent approach, equation~\eqref{eq:eom0} must be solved for the fluid and magnetic variables simultaneously. For barotropic fluids (where $\nabla p \times \nabla \rho = 0$), we follow \cite{rox66} by dividing both sides by the density and taking the curl, leading to
\begin{equation} \label{eq:truegs}
0 = \nabla \times \left\{ \frac{\nabla\psi}{\rho r^2 \sin^2\theta} \left[ \tilde{\Delta} \psi + \frac{I(\psi) I'(\psi)}{L'(F^2)} \right]  \right\},
\end{equation}
which can be used to solve for the poloidal flux function if the density is known. The density will not, in general, be spherically symmetric unless the field is force-free, which makes finding solutions difficult even in the Maxwell case \cite[see, e.g.,][]{hask08}. Even in a magnetar, however, the magnetic energy density is expected to be small relative to that of the hydrostatics, with equipartition defined by the virial limit $B_{\rm virial} \approx 1.4 \times 10^{18} (\Ms/1.4 M_{\odot}) (R_{\star}/10\text{ km})^{-2}$~G for stellar mass $M_{\star}$. Thus, so long as the internal field $B \ll B_{\rm virial}$, a perturbative approach can be reliably used, where the magnetic field is superimposed on a spherical background. Writing $X \to X + \delta X$ for \emph{fluid} variables $X$, the perturbed version of equation \eqref{eq:poloidaleq} reads
\begin{equation} \label{eq:perturbed}
\nabla \delta p + \delta \rho \nabla \Phi + \rho \nabla  \delta \Phi = \frac{\nabla\psi}{4 \pi r^2 \sin^2\theta} \left[ \tilde{\Delta} \psi + \frac{I(\psi) I'(\psi)}{L'(F^2)}  \right],
\end{equation}
which we henceforth simplify further by taking the Cowling approximation ($\delta \Phi = 0$). Such an approximation is expected to lead to a factor $\lesssim 2$ \emph{underestimate} for the mass quadrupole moment of the star \citep{yosh13}; see Sec.~\ref{sec:quadellip}.

The system \eqref{eq:perturbed} can be closed by postulating an EOS relating the perturbed pressure and density, which need not necessarily match that of the background. A further simplification can be made through the non-barotropic approach devised by \cite{mast11}, where $\delta p$ and $\delta \rho$ are instead treated as \emph{independent} with the magnetic flux, $\psi$, being prescribed by hand to close the system. Following these authors, a field configuration is chosen such that: (i) the poloidal component is continuous with an external dipole field (i.e., expression \ref{eq:exact}), (ii) the current density vanishes on the surface, and (iii) the toroidal field resides within the neutral curves. The third of these requirements is automatically fufilled by our earlier choice \eqref{eq:toroidalchoice}.

\begin{figure*}
\includegraphics[width=\textwidth]{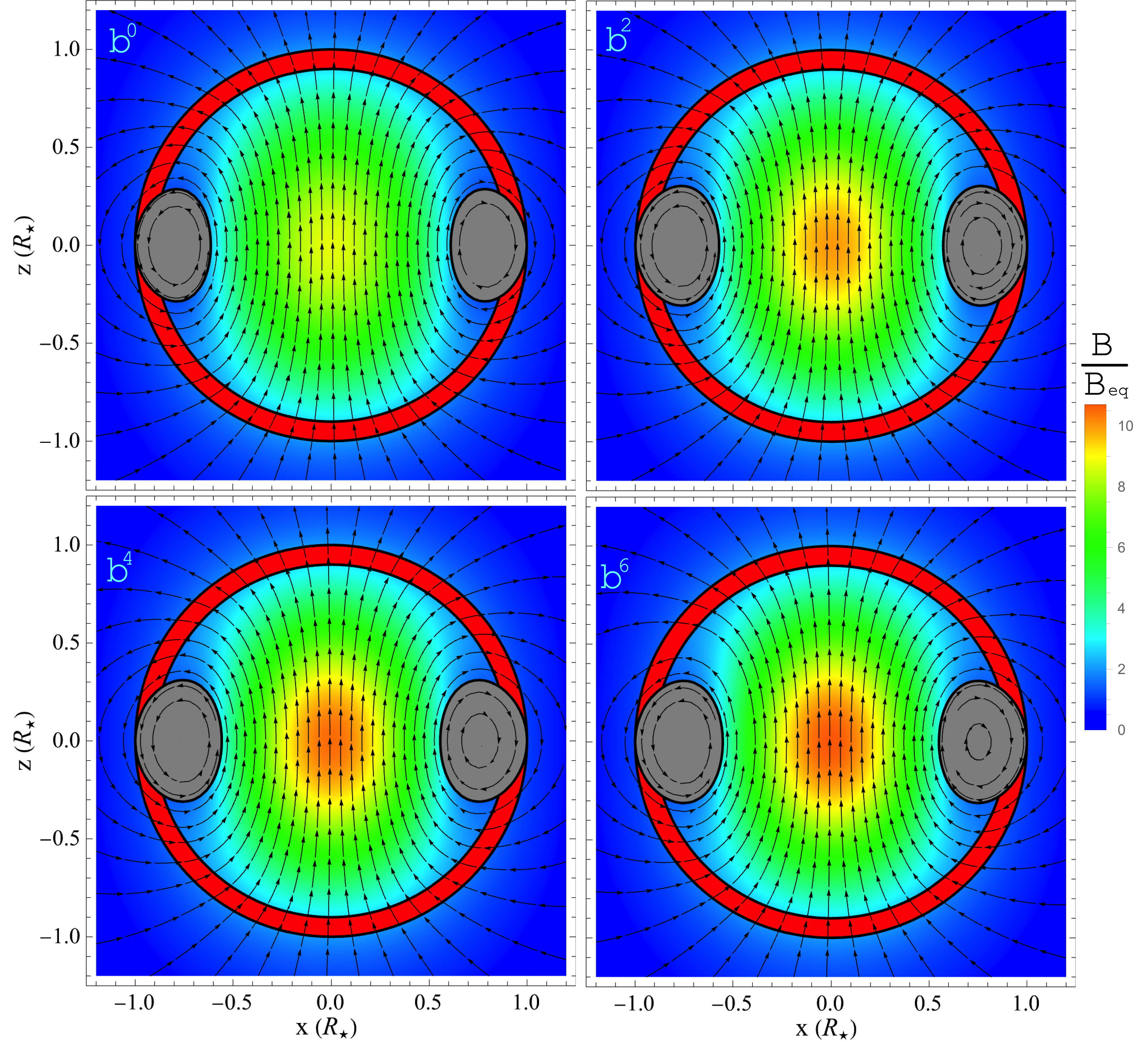}
\caption{Poloidal field lines associated with expression \eqref{eq:radialfunc} at successive orders in the BI parameter $b$, as per plot legends (increasing in PM order from top-left to bottom-right). The colour scale depicts the poloidal field strength (redder shades indicating greater $|\bb_{\rm pol}|$) relative to the equatorial value, $B_{\rm eq}$, for fixed $B_{\rm eq} = 7b/20$, $d_{0} = 0$, and $d_{i} = -1$ for $i=1,2,3$. The red annulus marks the location of a hypothetical crust, $0.9 \leq r \leq 1$. At the outer boundary, the field matches continuously to the perturbative PPM solution \eqref{eq:bio6}.
\label{fig:mags}}
\end{figure*}

In general, as the poloidal field has two components while the current is azimuthal at the surface, we require a functional form of $\psi$ involving at least three parameters to be fixed.  Following \cite{akg13}, we choose a polynomial of the form
\begin{equation} \label{eq:generalfield}
\begin{aligned}
\psi(r,\theta) &\equiv f(r) \sin^2 \theta \\
&= \frac{B_{p}}{2} \left( a_{2} r^2 + a_{4} r^4 + a_{6} r^6 + d r^8 \right) \sin^2 \theta ,
\end{aligned}
\end{equation}
where the minimum power ($r^{2}$) is chosen so that the current is everywhere finite. The extra constant $d \equiv a_{8}$ can be kept to explore a wider range of equilibria. In order to consistently match with an external field \eqref{eq:bio6} (or any other), we further write each (dimensionless) constant as a series in inverse powers of $b$, such as $a_{2} = a_{20} + a_{21}B_{p}^2/b^2 + a_{22} B_{p}^4 /b^4 + a_{23} B_{p}^6/b^6$, and equate coefficients at each order. We remark at this point that equation \eqref{eq:perturbed} is solved in subsequent sections with the full \emph{nonlinear} Lagrangian $L_{\rm BI}$ (as we may have $B \gtrsim b$ for $r \ll 1$), even though we match to the perturbative solution at the magnetospheric boundary (where $B \lesssim 0.7 b$; cf. Fig.~\ref{fig:lei}). Imposing conditions (i) and (ii) described above, we find the lengthy expression {(again we refer to Appendix~\ref{sec:appb})}
\begin{widetext}
\begin{align}\label{eq:radialfunc}
f(r) =& \frac{B_{p}}{2} \Bigg\{ \left[ \frac{35}{8} -d_{0} + \frac{\left(\tfrac{143}{480} - d_{1} \right) B_{p}^2}{b^2} + \frac{\left(\tfrac{10013}{672000} - d_{2}\right) B_{p}^4}{b^4} - \frac{\left(\tfrac{126017}{2257920}+d_{3}\right)B_{p}^6}{b^6} \right] r^2 \nonumber \\
&- \left[ \frac{21}{4} - 3 d_{0} + \frac{\left(\tfrac{39}{80}-3d_{1}\right) B_{p}^2}{b^2} + \frac{\left(\tfrac{589}{22400} - 3d_{2}\right) B_{p}^4}{b^4} - \frac{\left(\tfrac{5479}{53760} + 3d_{3}\right) B_{p}^6}{b^6} \right] r^4 \\
&+ \left[ \frac{15}{8} - 3 d_{0} + \frac{\left(\tfrac{33}{160}-3d_{1}\right) B_{p}^2}{b^2} + \frac{\left(\tfrac{527}{44800} - 3d_{2}\right) B_{p}^4}{b^4} - \frac{\left(\tfrac{126017}{2688000} + 3d_{3}\right) B_{p}^6}{b^6} \right] r^6 + \left( d_{0} + \frac{d_{1} B_{p}^2} {b^2} + \frac{d_{2} B_{p}^4} {b^4} + \frac{d_{3} B_{p}^6} {b^6} \right)r^8 \Bigg\}, \nonumber
\end{align}
\end{widetext}
which agrees with that found by \cite{akg13} as $b \to \infty$ for arbitrary $d_{0}$  (their A5).

Poloidal field lines for the flux function \eqref{eq:generalfield} are shown in Figure~\ref{fig:mags} at increasing orders in the expansion parameter $b$ [from top-left to bottom-right, $\mathcal{O}(b^{0},b^{2},b^{4},b^{6})$, as indicated in the figure legends]. We fix $B_{p} = 0.7 b$, $d_{0} = 0$, and $d_{i} = -1$ for $i=1,2,3$. The colour scale depicts the magnitude of the poloidal field relative to the equatorial value $B_{\rm eq} = 7 b/20$, while the toroidal region (whose magnitude depends on the unspecified parameter $i_{0}$) resides in the grey, shaded region. We see that the field matches continuously with respect to a (BI-augmented) dipole field at each order, while the core structure and field-strength maximum change somewhat between panels. The largest such change is with respect to the leading-order corrections, as expected, with the core field strength increasing by a factor $\sim 1.5$ between the Maxwell and $\mathcal{O}(b^6)$ cases. One feature of note is the widening of the toroidal domain in the BI cases compared to the Maxwell one. While this effect can be suppressed by tuning $d_{i}$ in principle, it is a consequence of the increased flux: $\psi > \psi_{c}$ in a larger volume. This implies that a given toroidal energy in the BI case corresponds to a locally weaker field since the volume increases. 

Although the above configurations are in some sense arbitrary because they have not been derived from the true equations of motion \eqref{eq:truegs}, they are qualitatively similar to cases which are. This is demonstrated in Appendix~\ref{sec:appa} showing the `true' solution for a constant-density star, where the leading-order ($b^{2}$) corrections lead to a similar increase in $B_{\rm max}$ and toroidal widening.

\subsection{Background hydrostatics} \label{sec:hydrostatics}

In the non-magnetic limit, suppose the spherical star is in hydrostatic equilibrium. For simplicity, we consider a background density profile akin to the Tolman VII solution,
\begin{equation} 
\rho =\rho_c (1-r^2),
\end{equation}
for central density $\rho_c=15 \Ms /(8\pi R_{\star}^3)$. The gravitational acceleration, found from the Poisson equation \eqref{eq:poisson}, reads
\begin{equation} 
\frac{d\Phi}{dr} = \frac{G M_{\star}}{2R_{\star}^2} r(5-3r^2),
\end{equation}
from which we reverse-engineer the EOS from the Euler equation \eqref{eq:eom0} with $\bb=0$ as
\begin{equation}
p(\rho) = p_{c} \frac{\rho^2 \left( \rho + \rho_{c}\right)}{2 \rho_{c}^3},
\end{equation}
where $p_{c} = 15 G \Ms^2 / (16 \pi R_{\star}^2)$ is the central pressure. While not motivated by any particular physical considerations, the Tolman VII profile matches reasonably well to realistic EOS and having an analytic profile simplifies subsequent calculations considerably; see, for instance, \cite{suvg23} and references therein for a discussion.

\section{Quadrupolar ellipticity} \label{sec:quadellip}

Equipped with the hydrostatic relations of Sec.~\ref{sec:hydrostatics}, we turn to solving the perturbation problem \eqref{eq:perturbed}. In fact, we do not need to solve this equation in its entirety as our primary interest relates to GWs, where only the density enters into the Newtonian mass quadrupole moment, $Q = \epsilon I_{0}$ for ellipticity $\epsilon$ and moment of inertia $I_{0}$, defining the GW strain \citep{thorne80}. Taking the curl of equation \eqref{eq:perturbed} gives a simple relation, at least in the Cowling approximation, for $\delta \rho$ which equation can be integrated up to an arbitrary function of radius which does not contribute to the ellipticity \citep{mast13},
\begin{equation}
\frac{ \partial \delta \rho}{\partial \theta} = - \frac{r}{R_{\star}} \frac{dr}{d\Phi} \left(\nabla \times \boldsymbol{f}_{\rm mag}\right)|_{\phi}.
\end{equation}

\begin{figure}
\includegraphics[width=0.487\textwidth]{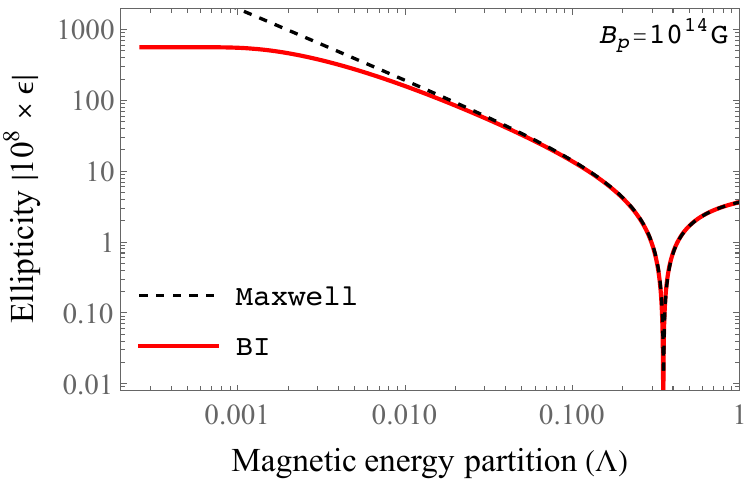}
\caption{Quadrupolar ellipticity \eqref{eq:ellip} as a function of the magnetic energy partition $\Lambda$ \eqref{eq:partition} for the Maxwell (dashed, black) and BI (solid, red) magnetic configurations \eqref{eq:radialfunc} for $B_{p} = 10^{14}$~G. For either theory, the ellipticity vanishes at $\Lambda = \Lambda_{\rm c} \approx 0.351$, with poloidally-dominated ($\Lambda > \Lambda_{\rm c}$) and toroidally-dominated  ($\Lambda < \Lambda_{\rm c}$) configurations corresponding to oblate ($\epsilon > 0$)  and prolate ($\epsilon < 0$) stars, respectively. The minimum partition, such that the BI toroidal function \eqref{eq:toroidalbi} is real, is $\Lambda_{\rm min} \approx 2.5 \times 10^{-4}$.
\label{fig:ellip14}}
\end{figure}

We define the ellipticity, $\epsilon$, according to \citep{mast11}
\begin{equation} \label{eq:ellip0}
\epsilon = \frac{I_{zz}-I_{xx}}{I_0},
\end{equation}
where $I_{jk}$ are the components of the moment-of-inertia tensor
\begin{equation}
I_{jk}=R_{\star}^5 \int_V dV \left(\rho + \delta\rho \right) \left(r^2\delta_{jk}-x_jx_k \right).
\end{equation}
We therefore have, in axial symmetry,
\begin{equation} \label{eq:ellip}
\epsilon = \frac{\pi R_{\star}^5}{I_{0}} \int^{1}_{0} dr \int^{\pi}_{0} d\theta \delta\rho(r,\theta) r^4\sin\theta(1-3\cos^2\theta).
\end{equation}
To compare with previous work, we introduce the poloidal-to-total magnetic energy ratio,
\begin{equation} \label{eq:partition}
\Lambda = \frac{\int_{V} dV \boldsymbol{B}_{p}^2/8\pi} {\int_{V} dV \bb^2/8\pi},
\end{equation}
which is a function of the parameter $i_{0}$ introduced in expression \eqref{eq:toroidalchoice}. The ratio $\Lambda$ is also a function of $B_{p}$ in general, as the problem is no longer scale-invariant in NLED. In practice, we define a grid of $i_{0}$ values for a given $\psi$, integrate expression \eqref{eq:ellip} after finding $\delta \rho$ for each $i_{0}$, and then infer the overall ellipticity profile as a function of the energy partition $\Lambda$. Note that $\Lambda = 1$ corresponds to a purely poloidal field while $\Lambda \ll 1$ is toroidally-dominated. Since the toroidal (poloidal) field generally leads to prolate (oblate\footnote{{Although the models constructed here are static, rotation would also induce some degree of geometric oblateness. While such a contribution is necessarily aligned with the symmetry axis and thus does not result in GW emission (which forms our primary interest here), some care is needed in applying this terminology in more general cases \cite[see, e.g.,][]{fr12}.}}) distortions, the limitation \eqref{eq:upperbound} implies a minimum $\Lambda$, $\Lambda_{\rm min}$, and a maximum prolateness of order ($b \gg B_{p}$)
\begin{equation} \label{eq:promax}
\epsilon_{\rm max}^{\rm prolate} \approx -6 \times 10^{-6} \left(\frac{b}{10^{16} \text{ G}}\right)^2 \left(\frac{1.4 M_{\odot}} {\Ms} \right)^{2} \left(\frac{\Rs} {10 \text{ km}}\right)^{4},
\end{equation}
as explored in more detail below. Note that results are not very sensitive to the choice of $\sigma$ made in \eqref{eq:toroidalchoice}.

Fixing $\Ms = 1.4 M_{\odot}$ and $\Rs = 10$~km, results with $d_{i} = 0$ are shown in Figure~\ref{fig:ellip14} for $B_{p} = 10^{14}$~G and $b=10^{16}$~G. The dashed curve shows the Maxwell result, which is in excellent agreement with that found by \cite{mast13}. The ellipticity drops sharply at $\Lambda = \Lambda_{\rm c} \approx 0.351$, below which the star is prolate ($\epsilon <0)$ rather than oblate ($\epsilon > 0$) as the toroidal field begins to dominate the deformation. For $\Lambda \gtrsim 10^{-2}$, the BI (solid) and Maxwell curves overlap, expected as $B_{p} = 10^{-2} b$. However, at extreme partitions where the toroidal component houses $> 99\%$ of the total magnetic energy, the classical ellipticity exceeds that of the BI prediction. This occurs because, although the field maximum increases relative to the Maxwell case  (see Fig.~\ref{fig:mags}), the overall deformation is reduced as $|L_{\rm BI}'(F^2)| < |L_{\rm Max}'(F^2)|=1$ and $I(\psi) \propto L'(F^2)$ (expression~\ref{eq:toroidaleq}).

A second important feature visible in Fig.~\ref{fig:ellip14} concerns minimum values of $\Lambda$. In the Maxwell case, this parameter can be arbitrarily\footnote{In practice, there is a limiting partition strength set by stability; \cite{akg13} argue $B_{\rm tor,max} \lesssim 10^{17} \text{ G} \times \left(B_{\rm p}/10^{15}\text{ G}\right)^{1/2}$ in the Maxwell case. When considering leading-order QED effects, \cite{rw21} find that magnetosonic-like instabilities, associated with Landau quantisation of fermions, can operate  on $\sim$~ms timescales for super-Schwinger fields, complicating the picture.} negative, suggesting that magnetar ellipticites could, in principle, be very large. In BI theory, because the toroidal field is limited by condition \eqref{eq:upperbound}, there is a maximum toroidal energy which depends on the characteristic strength $B_{p}$. For $B_{p} = 10^{14}$~G we find $\Lambda_{\rm min} \approx 2.5 \times 10^{-4}$, which limits the maximum magnitude of the ellipticity to $|\epsilon_{\rm max}| \approx 5.7 \times 10^{-6}$, which has implications for GW  detectability (Sec.~\ref{sec:gwdetect}) and other phenomena (Sec.~\ref{sec:magphenom}). Note the curve almost plateaus for $\Lambda \lesssim 10^{-3}$.

\begin{figure}
\includegraphics[width=0.487\textwidth]{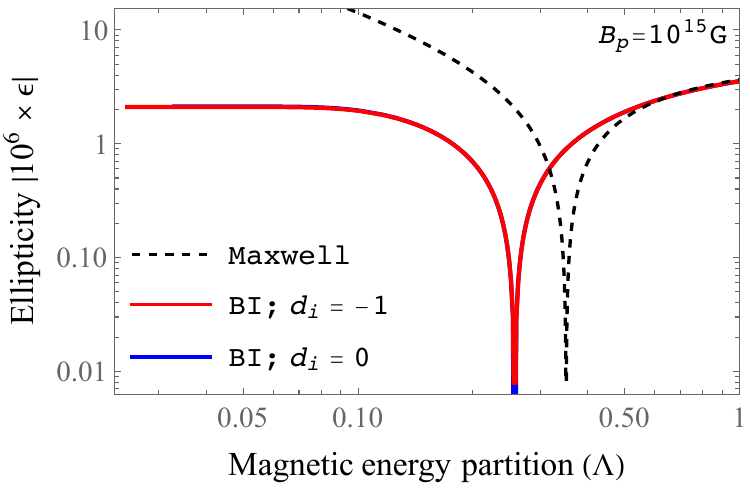}
\caption{Similar to Fig.~\ref{fig:ellip14} except for $B_{p} = 10^{15}$~G and different values for the free parameters $d_{i}$ (see plot legends). The critical energy ratio for the BI case is $\Lambda_{\rm BI,c} \approx 0.258$, with $\Lambda < \Lambda_{\rm BI,c}$ corresponding to prolate stars. We have $\Lambda_{\rm min} \approx 2.5 \times 10^{-2}$ for the case with $d_{i} = -1$ (red) and $\Lambda_{\rm min} \approx 3.3 \times 10^{-2}$ for $d_{i} = 0$ (blue). The latter curves overlap throughout most of the domain.
\label{fig:ellip15}}
\end{figure}

Figure~\ref{fig:ellip15} shows instead a case with $B_{p} = 10^{15}$~G, where the field is strong enough throughout the star such that the NLED impact is more evident across the whole range of energy partitions. For comparison, we also show ellipticites for a case where the $d_{i}$ are not zero (red curve), demonstrating that the results are visually indistinguishable. For this value of $B_{p}$, the quadrupole moment is smaller than in the Maxwell case by $\approx 3\%$ for a purely poloidal field $(\Lambda = 1)$. This occurs again because the Lagrangian function $L'(F^2)$ entering the GGS operator is smaller in magnitude, though the relative decrease is not negligible like in the case of $B_{p} = 10^{14}$~G (Fig.~\ref{fig:ellip14}). The partition for which the poloidal oblateness and toroidal prolateness balance reads $\Lambda_{\rm c} \approx 0.258$. Such a shift reflects the fact that the poloidal energy in the core increases relative to the Maxwell case (Fig.~\ref{fig:mags}). 

Because the local, poloidal field approaches $b$, the toroidal sector is much more restricted in this case, with $\Lambda_{\rm min} \gtrsim 10^{-2}$ depending on the choice for the $d_{i}$. Larger, negative values of the $d_{i}$ effectively increase the toroidal volume (see Fig.~\ref{fig:mags}) which means that, for a given energy, the maximum $I(\psi)$ attained there is lower, and hence a smaller value of $\Lambda_{\rm min}$ is permitted. Either way, such minima strongly limit the ellipticity range, with the \emph{maximum} prolateness being a factor few smaller than the purely-poloidal oblateness. If we were to increase $B_{p}$ further, progressively smaller values for $i_{0}$ would be demanded to prevent $T(r,\theta)$ becoming imaginary, with only oblate stars being permitted for $B_{p} \gtrsim 2 \times 10^{15}$~G (see Sec.~\ref{sec:gwdetect}).

\subsection{Gravitational-wave detectability} \label{sec:gwdetect}

The GW-detectability of a magnetar depends primarily on its polar field strength, birth period, and internal geometry as the surface field together with the ellipticity control the spindown rate. For a biaxial star, the GW frequency is twice the spin frequency, $f_{\rm GW} = 2 f$, and $\epsilon$ features directly in the expression for the characteristic GW strain \cite[e.g.,][]{aasi14}
\begin{equation} \label{eq:h0}
h_{0} = \frac{16 \pi^2 G}{c^4} \frac{I_{0} |\epsilon| f^2}{D},
\end{equation}
for source distance $D$. A maximum value for $|\epsilon|$ due to the toroidal limitation \eqref{eq:upperbound} implies a maximum GW signal-to-noise ratio (SNR) for a given initial $f$, $D$, and so on. Under the optimistic assumption of optimal matched-filtering, the SNR, for an object that is tracked for a time $\tau$ where $f_{\rm GW}(\tau) = f_{f}/2$ but initially $f = f_{i}/2$, reads \cite[e.g.,][]{dall09}
\begin{equation} \label{eq:snrform}
\text{SNR} = 2 \left[ \int^{f_{f}}_{f_{i}} df \frac{|\tilde{h}(f)|^2}{S_{n}(f)}, \right]^{1/2} 
\end{equation}
where $|\tilde{h}(f)|^2 \approx \tfrac{1}{2}h^2[f(t)]/|\dot{f}_{\rm GW}|$ in the stationary-phase approximation and $S_n$ is the noise spectral density of the detector.

\begin{table*} \label{tab:snrtab}
 \centering
  \caption{Comparison of quadrupole-related quantities for deformed magnetars in the BI theory, together with the distance corresponding to unit SNR for GW-detectability with either DECIGO or ET using expression \eqref{eq:snrform}, for a given field strength $B_{p}$ assuming the maximum deformation $|\epsilon| = \text{max}\{|\epsilon^{\rm pol}|,|\epsilon^{\rm tor}|\}$. We fix $b=10^{16}$~G, $d_{i} = 0$, $\Ms = 1.4M_{\odot}$, $R_{\star}=10$~km, $f_{i} = 10$~Hz, and $f_{f} = 1$~Hz.}
  \begin{tabular}{lcccccc}
  \hline
    \hline
     $B_{p}$~(G) & $\Lambda_{\rm min}$& $\epsilon^{\rm pol}(\Lambda = 1)$& $\epsilon^{\rm tor}(\Lambda = \Lambda_{\rm min})$ & $D_{\rm DECIGO}(\text{SNR}=1)$~(kpc) & $D_{\rm ET}(\text{SNR}=1)$~(kpc) & $\text{SNR}\left(\frac{\text{Maxwell}}{\text{NLED}}\right)$ \\
\hline
$10^{14}$ & $2.51 \times 10^{-4}$  & $3.65 \times 10^{-8}$  & $-5.64 \times 10^{-6}$ & 86.6 & 21.8 & 14.0 \\
$5 \times 10^{14}$ & $7.76 \times 10^{-3}$  & $9.07 \times 10^{-7}$ & $-4.77 \times 10^{-6}$  & 14.7 & 3.69 & 13.1 \\
$10^{15}$ & $3.29 \times 10^{-2}$ & $3.55 \times 10^{-6}$ & $-2.13 \times 10^{-6}$ & 5.45 & 1.37 & 15.3 \\
$2 \times 10^{15}$ & $9.96 \times 10^{-2}$ & $1.32 \times 10^{-5}$ & $7.49 \times 10^{-6}$ & 10.1 & 2.55 & 8.6 \\
$4 \times 10^{15}$ & $2.86 \times 10^{-1}$ & $4.35 \times 10^{-5}$ & $3.78 \times 10^{-5}$ & 16.7 & 4.20 & 5.4 \\
\hline
  \hline
\end{tabular}
\end{table*}

Although often applied from a birth frequency of $\sim 1$~kHz, using only the strain \eqref{eq:h0} in expression \eqref{eq:snrform} is not really appropriate in that case because the signal may be contaminated by GWs from the non-axisymmetric core collapse and quasi-normal ringing. Supposing instead the star is tracked from a magnetar-like frequency $f_{i} = 10$~Hz until $f_{f} = 1$~Hz (corresponding to $\gtrsim 4$~yr of spindown for $B_{p} \approx 4 \times 10^{15}$~G), we estimate the distance such that the SNR \eqref{eq:snrform} equals one using the noise curves from figure 2 of \cite{yagi13}. The best-suited detector for this band is the planned DECi-hertz Interferometer GW Observatory (DECIGO) for which we estimate $S_{n} \approx 10^{-49} (f/1\text{ Hz})^{2} \text{ Hz}^{-1}$ around $f \gtrsim 1$~Hz by fitting data from \cite{yagi13}. In this band, the Einstein Telescope (ET) sensitivty is instead $S_{n} \approx 1.1 \times 10^{-43} (f/1\text{ Hz})^{-4} \text{ Hz}^{-1}$. Note the different scalings with frequency: DECIGO (ET) improves (worsens) in sensitivty as $f_{i} \to f_{f}$. We compute the SNR assuming spindown appropriate for an orthogonal rotator, so that $\dot{f} \propto K_{d} f^3 + K_{\rm GW} f^5$ for dipolar and GW contributions with appropriate constants $K$ \citep{dall09}. Note that we ignore NLED augmentations to spindown, which are weak unless $B \sim b$; see Sec.~\ref{sec:spindown}.

For canonical values $\Ms=1.4 M_{\odot}$ and $R_{\star}=10$~km, the results are shown in the fifth and sixth columns of Table~\ref{tab:snrtab} for DECIGO and ET, respectively. For reference, we also list the minimum permitted $\Lambda$ such that \eqref{eq:upperbound} is respected, together with the ellipticities calculated from equation \eqref{eq:ellip} in either the maximally-toroidal or purely poloidal states, and the relative SNR between Maxwell and NLED cases for $\Lambda_{\rm min}\leq\Lambda\leq1$. Note in particular that once $B_{p} \gtrsim 0.2 b = 2 \times 10^{15}$~G, restriction \eqref{eq:upperbound} becomes so severe that the star can \emph{only be oblate}. A consequence of this is a turnover in the distance formula, so that adolescent magnetars with $B_{p} \gtrsim 10^{15}$~G may be more difficult to detect than previously estimated. Still, detections out to the Magellanic clouds are feasible with DECIGO: for typical strengths $B_{p} \sim \text{few} \times 10^{14}$~G, the detection horizon is $\lesssim 50$~kpc.

\section{Implications for magnetar phenomena} \label{sec:magphenom}

We turn to exploring some ways in which the constructions and estimates made in previous sections may manifest in astrophysical observations involving magnetars. These include that relevant for crustal field evolution (Sec.~\ref{sec:hallohm}), spindown (Sec.~\ref{sec:spindown}), GRB phenomena (Sec.~\ref{sec:grbs}), overstraining events and flares (Sec.~\ref{sec:failures}), and (anti)-glitches (Sec.~\ref{sec:glitches}).

\subsection{Hall-Ohm evolution} \label{sec:hallohm}

How does the magnetic field in the crust of a neutron star evolve in NLED? The physical picture we adopt is the standard one involving a stationary lattice of ions strewn with mobile electrons whose motion gradually advects the field lines, while the finite conductivity of the material leads to Ohmic dissipation. The evolution of the \emph{core} field will also be affected by changes to the effective current \eqref{eq:ampere}, though the topic of core evolution (a l{\'a} ambipolar diffusion) even for the Maxwell case is not fully agreed upon \cite[e.g.,][]{pass17,gus17}. As the Lorentz force has the same functional form in BI-MHD \eqref{eq:lorentz}, the particle momentum equations remain the same and it is only the Maxwell sector that differs. In considering non-ideal evolution, the electric field should strictly-speaking be restored within the Lagrangian function, viz. $F^2 = \tfrac{1}{2}(\bb^2 - \bE^2)$, and perhaps the dual $G^2 \neq 0$. However, as discussed in Sec.~\ref{sec:mhd}, we operate within the MHD regime such that $\bE \sim (v/c) \bb \ll \bb$ and hence these contributions are negligible except in cases with unphysically large couplings.

To derive an equation for the crustal field evolution, we note that the NLED Faraday equation persists in the usual Maxwell form \cite[e.g.,][]{boil70,kim00,per18},
\begin{equation} \label{eq:faradeqn}
\frac{\partial \bb}{\partial t} = - c \nabla \times \bE.
\end{equation}
Using Ohm's law\footnote{Since the particle momentum equations are the same as in Maxwellian MHD, this law can be `derived' using the standard arguments given by \cite{gr92}; see equation (13) therein, where we drop chemical terms assuming fluid motions occur instantaneously relative to magnetic evolution timescales. Thermoelectric terms and those accounting for electron inertia are similarly dropped.},
\begin{equation} \label{eq:ohm}
\bE = \bj/\sigma_{0} + \bj \times \bb / n_{e} e c ,
\end{equation}
where we make use of the definition $\bj = -n_{e} e \boldsymbol{v}$, and taking its curl leads to the familiar induction equation from \eqref{eq:faradeqn}
\begin{equation} \label{eq:ind0}
\frac{\partial \bb}{\partial t} = -\nabla \times \left ( \frac{c \bj}{\sigma_{0}} + \frac{\bj \times \bb}{n_{e} e} \right).
\end{equation}
In the above we have introduced the electrical conductivity parallel to the magnetic field, $\sigma_{0}$, and electron number density $n_{e}$. In a strongly-magnetised medium, the conductivity effectively bifurcates into components which are parallel and orthogonal to $\bb$; this is treated consistently in the non-quantising limit as above through the Hall term \cite[see][]{urp80}. Quantum effects complicate the picture significantly, as $\sigma_{0}$ can change by factors of $\sim 10$ due to Shubnikov-de-Haas oscillations \citep{rw23}. A strong magnetic field may also influence the background EOS: effects involving spin polarisation and Landau quantisation can adjust the bulk properties of the star, with an overall stiffening for $|\bb| \gg B_{\rm QED}$ \citep{card01}. A magnetic stratification could also lead to battery-like effects implicitly ignored in equation \eqref{eq:ind0}.

Incorporating the NLED Amp{\'e}re law \eqref{eq:ampere}, we arrive at the generalised Hall-MHD induction equation
\begin{equation} \label{eq:indf}
\frac{\partial \bb}{\partial t} = \nabla \times \left\{ \frac{c^2 \nabla \times \left[ L'(\tfrac{1}{2}\bb^2) \bb \right]}{4 \pi \sigma_{0}} + \frac{c \nabla \times \left[ L'(\tfrac{1}{2}\bb^2) \bb \right] \times \bb}{4 \pi n_{e} e} \right\}.
\end{equation}
One sees from equation~\eqref{eq:indf} that the Ohm and Hall timescales should \emph{increase} relative to their Maxwellian counterparts by a factor $\sim \sqrt{1+\bb^2/b^2}$. This implies that ultra-strong fields in magnetars may be preserved for longer, though the effects are minimal unless $\bb^2 \gtrsim b^2$. An interesting aspect of \eqref{eq:indf} is that, even in the absence of the Hall term, multipolar components should decay in unison to some degree due to the nonlinear coupling. In the Maxwell case, the decay of Ohmic modes are decoupled meaning that one may expect the dipole field to be dominant at late times as it decays slowest. 

{Tackling equation \eqref{eq:indf} in a fully-numerical setup is challenging owing to the fact that even the most stable magnetic-evolution codes are reliable only at local strengths $\lesssim 10^{16}$~G, as the Hall prefactor grows large \cite[see, e.g.,][]{pv19}, where NLED effects begin to dominate. Moreover, without an analytic form describing a potential field, an elliptic subroutine must also be devised to sequentially feed boundary data to ensure consistent matchings at $r=1$. This is reminiscent of evolution schema involving couplings between the crust and magnetosphere where currents thread the surface, as described (for instance) by \cite{urban23}; techniques presented there could be useful in this direction \cite[especially via neural networks for rapid matching; see][]{urb25}. While a special-purpose code could be built using various methods for handling stiffness (e.g. Radau), this lies beyond the scope of the present article.}

We provide a simple, worked example for field evolution using the `volume-averaged' scheme introduced by \cite{ag08}, where spatial dependencies are washed out by treating $\bb$ as a scalar and replacing $\nabla \rightarrow 1/\tilde{L}$ for system length-scale $\tilde{L}$. For a crust-confined dipole, we expect $\tilde{L} \sim 0.1 R_{\star}$. In this scheme, equation \eqref{eq:indf} becomes
\begin{equation} \label{eq:volumeaveraged}
0 = \dot{B}(t) + \frac{c}{4 \pi \tilde{L}^2 \sqrt{1+ \frac{B(t)^2}{b^2}}} \left[ \frac{c B(t)}{ \sigma_{0}} + \frac{B(t)^2}{n_{e} e} \right],
\end{equation}
which can be easily solved subject to the initial condition $B(0) = B_{0}$ for initial (volume-averaged) field strength $B_{0}$.

\begin{figure*}
\includegraphics[width=\textwidth]{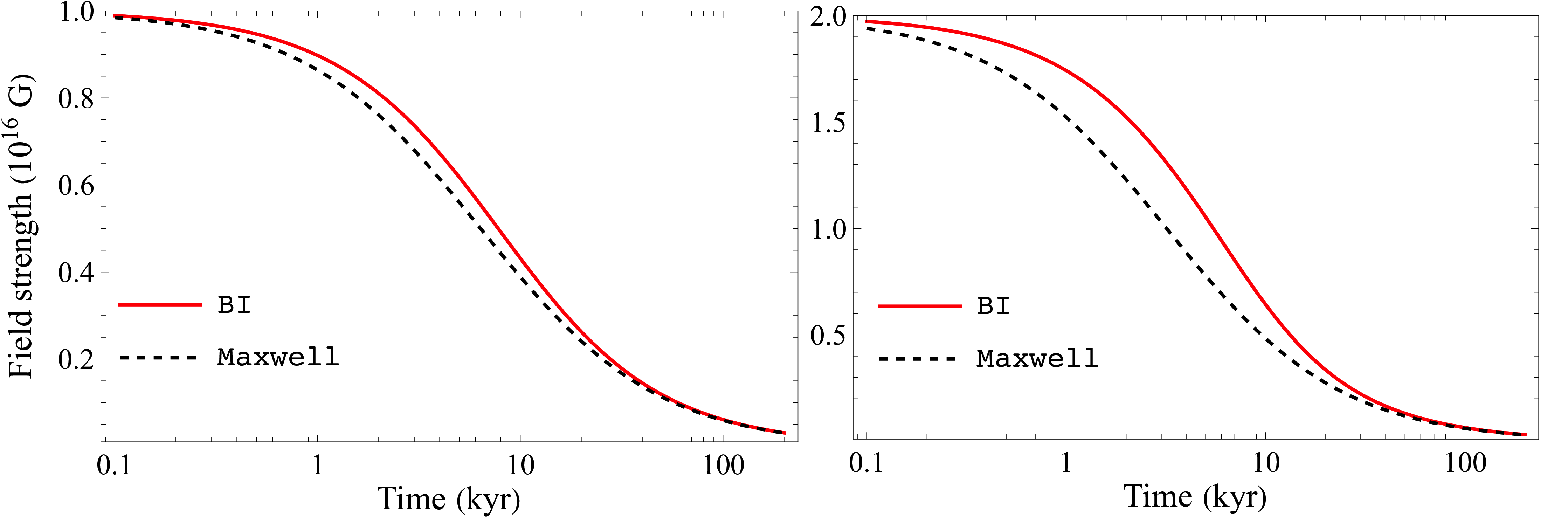}
\caption{Solutions to equation \eqref{eq:volumeaveraged}, describing the volume-averaged field evolution as a function of time in BI theory (solid curves) relative to the Maxwell case (dashed), subject to the initial condition $B_{0} = b = 10^{16}$~G (left) or $B_{0} = 2 \times 10^{16}$~G (right).
\label{fig:hallev}}
\end{figure*}

Figure~\ref{fig:hallev} shows solutions to equation~\eqref{eq:volumeaveraged} for $B_{0} = b$ (left) and $B_{0} = 2 b$ (right) where we prescribe some representative parameters relevant for the inner crust, viz. $n_{e} = 10^{36} \text{ cm}^{-3}$ and $\sigma_{0} = 10^{24} \text{ s}^{-1}$. For contrast, we show both the BI (solid, red) and Maxwellian (dashed, black) tracks. As expected, the BI field is always larger than the Maxwellian one for $t>0$, though the effects are small unless $B \gtrsim b$. After a kyr of evolution we find, for example, $B_{\rm BI}/B_{\rm Max} \approx 1.14$ for $B_{0} = 2b$. At times $t \gtrsim 10^{2}$~kyr the curves are indistinguishable as $B(t) \ll b$ by this point irrespective of the initial field. The temporary preservation of a strong field, as depicted in Fig.~\ref{fig:hallev}, could influence the stellar evolutionary track in a few ways. 

Because magnetic decay is stalled, Joule heating is reduced in kind by a factor $\sim L'(F^2)$ implying that highly-magnetised objects should appear colder. This may help to explain why (some members of) the class of `ultra-long period' objects (ULPs) appear to be both cold and magnetar-like, assuming they are isolated neutron stars \cite[cf.][for a critical discussion]{ruit24}. These sources were seen to pulse in the radio band in a way that is phenomenologically similar to pulsars and yet, {with the exception of ASKAP J1832--0911 \citep{wang24}}, were invisible to follow-up X-ray searches \cite[e.g.,][]{hw22}. If they did indeed house polar field-strengths of order $\gtrsim 10^{16}$~G, as would fit with spindown bounds and pair-production thresholds \citep{suvm23}, alleviating tensions with thermal limits becomes easier. {Alternative pathways for ULP production that do not require NLED-level fields have been explored though (see \citealt{coop24,sdp25}.}



\subsection{Spindown} \label{sec:spindown}

Maintaining a stronger polar-field strength also implies that the star should spin-down \emph{faster} than in the Maxwellian case, where the spindown rate of an inclined rotator in vacuum reads \cite[e.g.,][]{man77}
\begin{equation} \label{eq:braking}
\dot{P}(t) =  4 \pi^2 \frac {B_{p}(t)^2 \sin^2 \chi(t) R_{\star}^{6}} {6 I_{0} P(t) c^3},
\end{equation}
for spin period $P$ and magnetic inclination angle $\chi$. In NLED, the Poynting flux, $\boldsymbol{S}$, is augmented by a factor which is related to the effective increase of the surface dipole, as per the perturbative solution \eqref{eq:exact} \citep{den16}. We have
\begin{equation}
\boldsymbol{S} = \frac{1}{4\pi} \bE \times \boldsymbol{H} = - \frac{L'(F^2)}{4\pi} \bE \times \bb, 
\end{equation}
which can be integrated given some magnetospheric solution to deduce the adjusted spindown rate. Such a calculation was carried out by \cite{heyl19} in the perturbative, PM framework finding an effective increase in the radiated power at the surface of the star in the form (see equations 2.22 and 3.10 therein)
\begin{equation}
\frac{L_{\rm PM}}{L_{\rm Max}} \approx 1+\frac{12}{5} \alpha_{\rm PM} B_{p}^2.
\end{equation}
For the BI theory, $\alpha_{\rm PM} = 1/ 2 b^2$ and hence for $B_{p} = 0.5 b$ spindown may be enhanced by $\approx 30\%$. However, care should be taken since the PM expansion becomes an increasingly poor approximation as the field grows: a counterterm $\sim B_{p}^4/b^4$ at PPM order will reduce the NLED contribution. Nevertheless, for $B_{p} \lesssim 0.5 b$ we anticipate a BI-adjusted spindown law of the form
\begin{equation} \label{eq:bispindown}
\dot{P}(t) \approx 4 \pi^2 \frac {B_{p}(t)^2 \sin^2 \chi(t) R_{\star}^{6}} {6 I_{0} P(t) c^3} \left[1 + \frac{6}{5} \frac{B_{p}(t)^2}{b^2} \right],
\end{equation}
which, combined with decay-stalling, could steepen the light curves of millisecond magnetars. Corrections to such a formula accounting for magnetospheric drag were studied by \cite{petri15}, who found the result $\sin^2 \chi \to k_{1} + k_{2} \sin^2\chi$ for $k_{1,2} \approx 1$.

\subsection{Gamma-ray bursts and dynamos} \label{sec:grbs}

In a neutron-star binary coalescence event where the constituents are not too heavy, a third, more extreme neutron star may be born. Due to (orbital) angular momentum conservation, such a remnant is expected to be rapidly rotating with a high degree of differential rotation that can excite dynamo activity \cite[e.g.,][]{kiuchi24}.

A hotly debated topic in the GRB literature concerns whether or not magnetar remnants are capable of launching a relativistic jet that can drill through the polar baryon pollution \cite[see][for a review]{ciolfi20}. Numerical simulations of mergers that leave a magnetar tend to find that magnetic field strengths approaching equipartition are needed to a launch a relativistic jet (if indeed a jet can be launched). However, if the toroidal field is capped due to condition \eqref{eq:upperbound}, this could set an effective bottleneck for both the hoop stress collimating the jet (`magnetic tower') and dynamo activity: the feedback cycle of poloidal-toroidal amplification halts as the toroidal field approaches the ceiling. If a magnetar were to be confirmed as a central engine in a merger-driven GRB (e.g. via pulsed emissions), the jet Lorentz factor may be used to implicitly set a lower bound on the BI critical field via the requirement $b \gtrsim B_{\phi}$. The unavailability of strong magnetic pressures may also reduce the collapse time of a supra- or hypermassive object \citep{suvg22}.

There is tentative evidence for $|\epsilon| \gg 10^{-6}$ from GRB afterglow data: \cite{lasky16} found that values of $|\epsilon| \sim 10^{-3}$ are consistent with some X-ray lightcurves \cite[such as GRB 090510; see also][]{suvk20}. Because of \eqref{eq:promax}, such an ellipticity can only be sourced by the poloidal field or non-magnetic mechanisms unless $b \gtrsim 10^{17}$~G (see Tab.~\ref{tab:snrtab}). Such values are competitive with terrestrial experiments \citep{akman18}. An oblate $\epsilon \sim 10^{-3}$ requires $B_{p} \approx \text{few} \times 10^{16}$~G, where NLED effects should be significant. More importantly, a poloidally-sourced ellipticity implies that dipolar spindown likely dominates over GW losses even for spin frequencies approaching breakup so that the braking index should not exceed $n \sim 3$ unless $b \gg B_{p}$, where $B_{p}$ could be large in a remnant.

\subsection{Crustal failures} \label{sec:failures}

A popular theory for outburst triggers in magnetars relates to \emph{crustal failures}. The basic picture is that once some critical stress threshold is met, a plastic deformation is initiated together with local magnetic relaxation such that field lines, anchored to the crust, are twisted \cite[e.g.,][]{xi16}. One of the key model ingredients is how elastic stresses compare to magnetic ones \citep{lg19}. A crude comparison can be made by simply comparing the magnitudes of the magnetic stress-tensor components between the Maxwell and NLED cases. To that end, consider the ratio (for each $i,j$) 
\begin{equation} \label{eq:Qratio}
\mathcal{R}(i,j) = \frac{L(F^2) \delta^{ij}- L'(F^2) B^{i} B^{j}}{ B^{i} B^{j} - \tfrac{1}{2}B^2 \delta^{ij}}.
\end{equation}
We calculate $\mathcal{R}$ for various tensorial components using the configurations detailed in Sec.~\ref{sec:magfield}, ignoring complications about the character of Lagrangian displacements induced by magnetic stresses \cite[see][]{koj24}.

\begin{figure*}
\includegraphics[width=\textwidth]{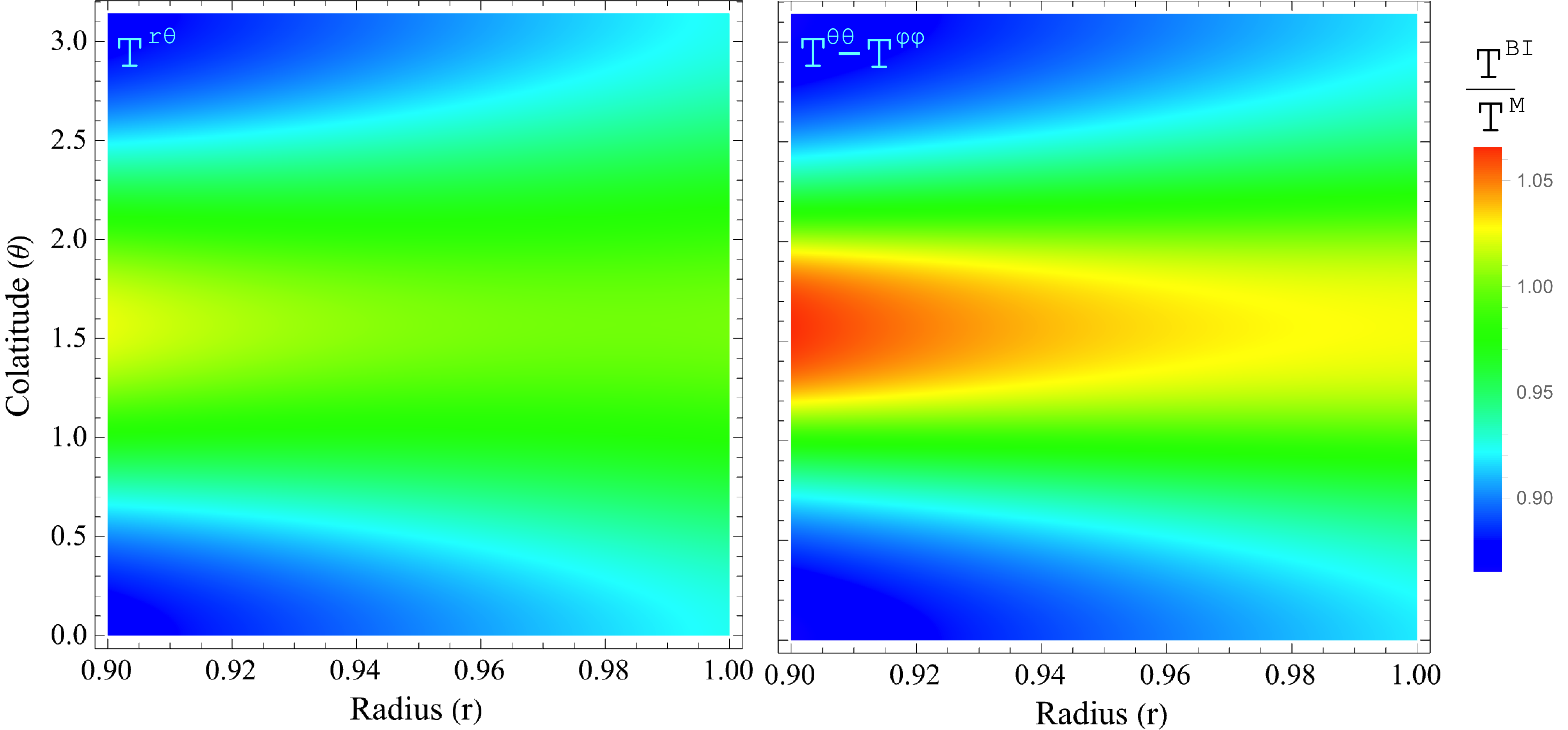}
\caption{Dimensionless ratios of BI-to-Maxwell stresses \eqref{eq:Qratio} in the $r\theta$ (left) and $\theta \theta - \phi \phi$ (right) directions within a `crust' for the field \eqref{eq:radialfunc}, with redder shades indicating a greater ratio $\mathcal{R}$. The classical polar strength is fixed as $B_{p} = b/2$.
\label{fig:tij}}
\end{figure*}

Two of the components identified as important by \cite{koj24} for the failure problem, namely those in the $r \theta$ (left) and $\theta \theta - \phi \phi$ (right) directions, are shown in Figure~\ref{fig:tij} for $B_{p} = 0.5b$. Attention has been restricted to the `crust' $0.9 \leq r \leq 1$ and, for simplicity, only poloidal stresses are shown ($i_{0} = 0$). The ratios span the range $0.87 \lesssim \mathcal{R} \lesssim 1.08$ with the lower and upper limits reached near polar $(\theta \approx 0,\pi)$ and equatorial ($\theta \approx \pi/2$) colatitudes, respectively, at the base of the crust $(r \approx 0.9)$. The fact that $\mathcal{R}$ exceeds unity near the equator implies one may expect more frequent failures there relative to the Maxwell case, with the reverse applying near the poles, since the elastic stress should counterbalance the electromagnetic stress. This is likely to skew the anticipated waiting time distribution for flares to later times, as Hall drift tends to tangle field lines near the pole and hence build stresses there \cite[see figure 2 in][]{perna11}. This will be investigated in future work.

\subsection{Magnetic (anti)-glitches} \label{sec:glitches}

As first described by \cite{i01}, rapid changes to the quadrupolar ellipticity of a magnetar may bring about glitch-like activity. By angular momentum conservation, a sudden reequilibriation $\epsilon_{\rm i} \rightarrow \epsilon_{\rm f}$ would lead to a fractional change in $f$ of order
\begin{equation} \label{eq:glitchfreq}
\frac{\delta f}{f} \approx \frac{2}{3} \left( \epsilon_{\rm i} - \epsilon_{\rm f} \right).
\end{equation}
While this model was discussed by \cite{grs15} in the context of the \emph{anti-glitch} seen in 1E 2259+586, a positive glitch is also possible if $\epsilon_{\rm i} > \epsilon_{\rm f}$ \citep{mast15b}. In the latter case, a dynamical decay (growth) in the poloidal (toroidal) sector could be expected. Such a rapid rearrangement would likely be accompanied by high-energy activity; for instance, crustal yielding could lead to a local relaxation in the magnetic field and hence a small change in $\epsilon$.

In the BI theory, maximum values for the toroidal ellipticity also imply an upper limit to the size of a magnetically-induced glitch. For SGR 1935+2154 with $B_{p} \approx 4 \times 10^{14}$~G, results from Tab.~\ref{tab:snrtab} indicate a \emph{maximum} shift of $\approx 5 \times 10^{-6}$ for $b = 10^{16}$~G. Using equation \eqref{eq:glitchfreq}, this implies a maximum glitch magnitude of $\delta f \approx (b/10^{16}\text{ G})^2$~$\mu$Hz for $P = 3.25$~s. This is about half the size of the large glitch seen in this object prior to the release of a fast radio burst in 2020 October \cite[$\delta f \approx 1.8^{+0.7}_{-0.5}$~$\mu$Hz;][]{youn23} suggesting a floor $b \geq \sqrt{2} \times 10^{16}$~G, assuming magnetic rearrangement was responsible for the burst and glitch. 

\section{Discussion} \label{sec:discussion}

In this work, we explore the impact of NLED on magnetar structure. After reviewing MHD basics (Sec.~\ref{sec:mhd}) coupled to some beyond-Maxwell theory (though focussing on BI;  Sec.~\ref{sec:bilag}), we solve for the leading-order dipole corrections to the force-free field in vacuum using a perturbative approach, finding expression \eqref{eq:exact} at the PM level in an arbitrary theory. The method introduced by \cite{mast11} is then used to match a non-linear interior field to a PPM vacuum (see Fig.~\ref{fig:mags}). We then compute the magnetically-induced biaxiality, quantified through the quadrupolar ellipticity \eqref{eq:ellip}, which leads to the emission of GWs as the star rotates. We find that the GW amplitude is generally \emph{reduced} in NLED owing primarily to restrictions on the toroidal field set by the azimuthal equation of motion \eqref{eq:toroidaleq}, which implies a maximum prolateness \eqref{eq:promax}. Such a maximum may be in tension with observations of long-period magnetar X-ray modulations made by \cite{mak14} and others if interpreted as free precession, or GRB afterglow data, as both seem to require $|\epsilon| \gtrsim 10^{-4}$, unless $b \gtrsim 10^{17}$~G. Even so, with DECIGO the prospects for detecting GWs from adolescent, Galactic magnetars are not overly pessimistic (see Tab.~\ref{tab:snrtab}).

We also explore how limitations on the toroidal partition together with the effective augmentation of the near-surface field impacts on magnetar phenomena. For example, we find that because the current is effectively reduced in NLED via the generalised Amp{\'e}re law \eqref{eq:ampere}, the effective Ohmic and Hall timescales are reduced slightly for super-Schwinger fields (see Fig.~\ref{fig:hallev}). This may help to explain the coldness and longevity of strong fields in the mysterious class of ULPs, assuming (a subset of them) are isolated magnetars \citep{hw22,suvm23,coop24}, especially since spindown is also accelerated for strong fields (see equation~\ref{eq:bispindown}). A thorough investigation of Hall-Ohm (and magnetothermal more generally) evolution in NLED, including crustal plasticity (Sec.~\ref{sec:failures}), is reserved for future work.

Several extensions of this study are natural. One path involves the inclusion of general-relativistic effects, which may have interesting interplays with NLED as the energy density $T^{tt}$ becomes large \citep{rez04,petri15}. It has been argued, for instance, that the Blandford-Znajek process is less effective in NLED \citep{li20b}, which could influence the properties of relativistic jets relevant for GRBs (Sec.~\ref{sec:grbs}). Explicit solutions to the GGS equation, either for non-barotropic or barotropic stars (compare the calculations in Secs.~\ref{sec:magfield} and Appendix~\ref{sec:appa}), could also be useful to study the stability of internal fields along the lines discussed by \cite{rw21} for the core or \cite{rw23} for the crust. It would also be worth tackling the magnetosphere problem in more detail. Using the methodology introduced by \cite{urban23} and \cite{stef23}, the force-free NLED structure of the magnetosphere could be solved for, including the possibility of rotation (`NLED pulsar equation'). The structure of coronal loops, for example, relating magnetic topology to flare properties and hotspots may differ because the effective twist is modulated by $L'(F^2)$, as per equation \eqref{eq:toroidaleq}. 

Finally, it is important to note we do not discuss superconductivity/fluidity in this paper, though an understanding of such phenomena is probably crucial for accurately modelling the stellar interior; it is unclear what role NLED may play in this case.

\section*{Acknowledgements}
{We thank the anonymous referee for constructive comments.} Support provided by the Conselleria d'Educaci{\'o}, Cultura, Universitats i Ocupaci{\'o} de la Generalitat Valenciana through Prometeo Project CIPROM/2022/13 is acknowledged. AGS extends his thanks to the members of the theoretical astrophysics group in T{\"u}bingen for the warm welcome he received while this work was being completed.


\section*{Data availability statement}
Observational data used in this paper are quoted from the cited works. No new data were produced here.



\appendix

\section{Radial prefactors in magnetic field expressions} \label{sec:appb}

In an effort to preserve the flow of text, details regarding derivations of a few lengthy expressions are presented here rather than the main body. 

Consider the force-free equation \eqref{eq:forcefree}, $0 = \tilde{\Delta} \psi$, where
\begin{equation} \label{eq:deltapsie}
\begin{aligned}
\tilde{\Delta} \psi =& \frac{\partial}{\partial r} \left[ L'\left(\frac{\nabla \psi \cdot \nabla \psi} {2 r^2 \sin^2 \theta}\right) \frac{\partial \psi}{\partial r} \right] \\
&+ \frac{\sin\theta}{r^2} \frac{\partial}{\partial \theta} \left[ \frac{L'\left(\frac{\nabla \psi \cdot \nabla \psi} {2 r^2 \sin^2 \theta}\right)}{\sin \theta} \frac{\partial \psi}{\partial \theta} \right] .
\end{aligned}
\end{equation}
Using the PPM Lagrangian function \eqref{eq:ppm}, $L(F^2) = -F^2 + \alpha_{\rm PM} F^4 + \alpha_{\rm PPM} F^6$, an approximate dipolar solution, $\psi(r,\theta) \approx f(r) \sin^2\theta$, can be found using a Legendre projection. Explicitly, inserting this form of $\psi$ into equation \eqref{eq:deltapsie}, multiplying the result by $d P_{1}(\cos\theta)/d\theta =  \cos \theta$, and then integrating over angles ($0 \leq \theta \leq \pi$) leads to the nonlinear, ordinary differential equation (omitting the implied radial dependence)
\begin{equation} \label{eq:ffg}
\begin{aligned}
0 =& \left( \frac{2f}{r^2} - f'' \right) + \frac{4}{5} \alpha_{\rm PM} \Big[ \frac{3 (f')^2 f''}{r^2} - \frac{2 (f')^3}{r^3}  \\
&+ \frac{f (f')^2 + f^2 f''}{r^4} - \frac{4 f^2 f'}{r^5} - \frac{6 f^3}{r^6} \Big] \\
&+ \frac{18}{35} \alpha_{\rm PPM} \Big[ \frac{5 (f')^4 f''}{r^4} - \frac{4 (f')^{5}}{r^5} \\
&+ \frac{2 f (f')^4 + 4 f^2 (f')^2 f''}{r^6} - \frac{8 f^2 (f')^3}{r^7} \\
&+ \frac{4 f^3 (f')^2 + 2f^4 f''}{r^8} - \frac{16 f^4 f'}{r^9} - \frac{20 f^5}{r^{10}} \Big],
\end{aligned}
\end{equation}
which has the expected solution $f(r) = B_{p}/2r$ for $\alpha_{\rm PM} = \alpha_{\rm PPM} = 0$. In general though, equation \eqref{eq:ffg} resists transform and other integration techniques and we cannot find a general solution. Instead, we seek a series expansion of the form
\begin{equation} \label{eq:fx}
f(r) = \sum_{j=0}^{J} \sum_{k=0}^{K} f_{j,k}(r) (\alpha_{\rm PM})^{j} (\alpha_{\rm PPM})^{k},
\end{equation}
for some functions $f_{j,k}$ and integers $J,K$, taking care to ensure dimensional consistency. It turns out that the functions $f_{j,k}$ \emph{can} be found exactly by matching orders. The case considered in the main text corresponds to $J = 2$ and $K=0$ (expression \ref{eq:exact}), defining the second-order PM expansion. For instance, the inhomogenous $\mathcal{O}(\alpha_{\rm PM})$ equation (sourced by Maxwell terms with $f_{0,0} = B_{p}/2r$) reads
\begin{equation}
0 = \frac{9 B_{p}^3}{10 r^9} + \frac{2 f_{1,0}(r)}{r^2} - f_{1,0}''(r),
\end{equation}
which has the particular solution
\begin{equation} \label{eq:leadingo}
f_{1,0}(r) = {B_{p}^3}/{60 r^7},
\end{equation}
which is precisely the second term appearing in expression \eqref{eq:exact} from the main text. Having found the solution \eqref{eq:leadingo}, as in other works \cite[e.g.][]{li20}, the $\mathcal{O}(\alpha_{\rm PM}^2)$ equation (sourced by first-order corrections) becomes
\begin{equation}
0 = \frac{129 B_{p}^5}{100 r^{15}} + \frac{2 f_{2,0}(r)}{r^2} - f_{2,0}''(r),
\end{equation}
which has the solution $f_{2,0}(r) = 43 B_{p}^5 / 6000 r^{13}$. This matches the last term in expression \eqref{eq:exact}. Note that these cumbersome numerical prefactors appear because of the factors of 2 in both the denominator within the argument of $L$ and the zeroth-order solution, together with the fact that we have opted not to normalise the PM parameters by factorials to avoid corrupting interpretations of the orders.

For the BI (or similar) theory where there is only one scale parameter, $b$, the double sum \eqref{eq:fx} collapses to 
\begin{equation} \label{eq:bksum}
f(r) = \sum_{j=0}^{J} f_{j}(r) b^{-2j},
\end{equation}
recalling that only even powers of $1/b$ appear in the PM expansion of the Born-Infeld action; see expression \eqref{eq:biexpansion}. Repeating the same exercise as above but in inverse powers of $b$ at one-higher order, $J=3$, one finds $f_{0} = B_{p}/2r, f_{2} = B_{p}^3/120 r^7$ (as above, recalling that $\alpha_{\rm PM} = 1/2b^2$) and
\begin{equation}
0 = \frac{93 B_{p}^4}{2800 r^{15}} + \frac{2 f_{4}(r)}{r^2} -  f_{4}''(r) = 0,
\end{equation}
with solution $f_{4}(r) = 31 B_{p}^5 / 168000 r^{13}$ and 
\begin{equation}
0 = -\frac{16437 B_{p}^6}{4 r^{21}} + \frac{2 f_{6}(r)} {r^2} - f_{6}''(r) = 0,
\end{equation}
with solution $f_{6}(r) = - 5479 B_{p}^7 / 14112000 r^{19}$. It is not difficult to recognise that the general solution takes the form $f_{j}(r) \propto B_{p}^{j+1} r^{-3j-1}$ with appropriate coefficients; the $\mathcal{O}(b^{-6})$ solution found above is that reported in expression \eqref{eq:bio6}. Though clumsy, this method can be extended to arbitrary values of $J, K$ (plus their extensions to PPPM and so on) and multipolarity to build perturbative, potential-field solutions in any given nonlinear theory of electrodynamics.

Deriving expression \eqref{eq:radialfunc} from the main text for the radial prefactor of the \emph{internal} field $(\psi_{I})$ follows the same order-matching logic. Through a slight abuse of notation we take 
\begin{equation}
\psi_{I} = f_{I}(r) \sin^2\theta = \sum_{i=2,4,6,8} \sum_{j=0,1,2,3} a_{ij} b^{-2j} r^{i} \sin^2\theta,
\end{equation}
for which we have \emph{sixteen} free constants $a_{ij}$. We match the above at the surface $(r=1)$ to expression \eqref{eq:bksum} describing the external solution $(\psi_{E})$ at each power of $b$: $\psi_{I}(1,\theta) - \psi_{E}(1,\theta) = 0+ \mathcal{O}(b^{-8})$, together with its first-order derivatives, $\left(\partial \psi_{I} / \partial r  - \partial \psi_{E} / \partial r\right)|_{r=1} = 0+ \mathcal{O}(b^{-8})$. This constrains eight of the $a_{ij}$. The current, found from expression \eqref{eq:ampere}, $\bj = - c \nabla \times \left[ L'(F^2) \bb \right]/4\pi$, can similarly be decomposed into inverse powers of $b$ (and/or multipoles). Noting that $\bj$ is only azimuthal at the surface because we match to a potential field, the condition $\bj_{\phi}|_{r=1} = 0+ \mathcal{O}(b^{-8})$ adds four constraints for a total of twelve. The matching problem thereby reduces to one of linear algebra, which can be straightforwardly handled with software-assisted Gaussian elimination. Though conceptually simple, the matrix equations are lengthy even for an Appendix. Still, carrying this out returns expression \eqref{eq:radialfunc} from the main text, noting there remains four free constants (labelled $d_{i}$ for $i=0,1,2,3$ in Sec.~\ref{sec:magfield}) because we impose twelve conditions. This was done to explore a wider range of theoretical equilibria.

\section{Hydromagnetic equilibria for constant-density stars} \label{sec:appa}

\begin{figure*}
\includegraphics[width=\textwidth]{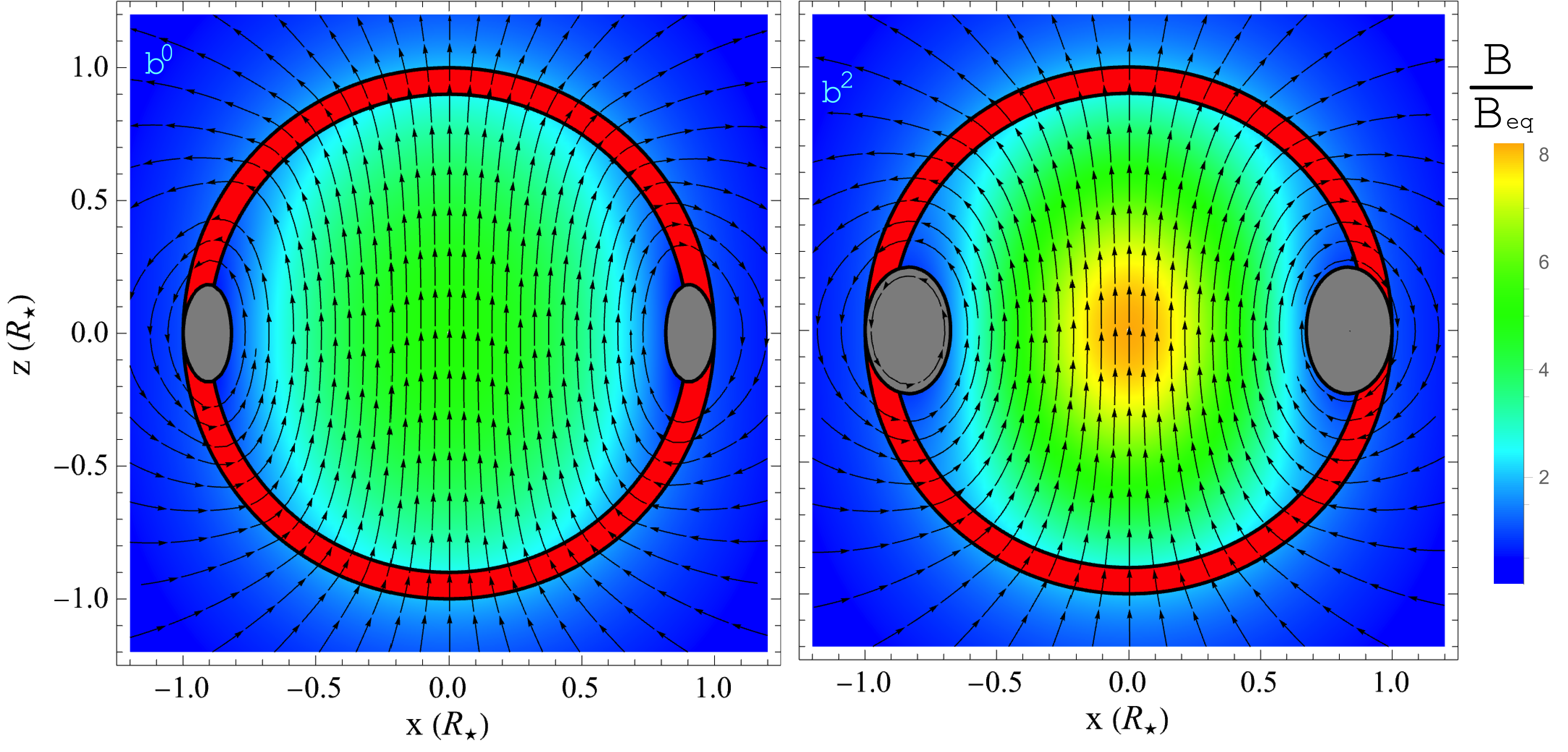}
\caption{Similar to Fig.~\ref{fig:mags}, though for solutions to equation~\eqref{eq:appeom} in the Maxwell (left) and BI (right; order $b^2$) cases for $B_{p} = 2b/3$.
\label{fig:cst0}}
\end{figure*}

Although the magnetic fields constructed in Sec.~\ref{sec:magfield} abide by a number of physically-motivated conditions, they do not arise as solutions to the equations of motion \eqref{eq:truegs}. To see how they compare with cases that \emph{do} solve the generalised \cite{rox66} equation, we consider the simple case of constant-density stars with dipolar, poloidal fields. 

For constant $\rho$ and $I(\psi) = 0$, equation \eqref{eq:truegs} becomes
\begin{equation} \label{eq:appeom}
\begin{aligned}
0 &= \nabla \times \left( \frac{\nabla\psi \tilde{\Delta} \psi}{r^2 \sin^2\theta}  \right) \\
&= \nabla \left(\frac{\tilde{\Delta} \psi}{r^2 \sin^2\theta} \right) \times \nabla \psi ,
\end{aligned}
\end{equation}
where $\tilde{\Delta}\psi$ is given in \eqref{eq:deltapsie}. It is clear that the set of magnetostatic equilibria in this case (and barotropic cases more generally) are restricted because the field must now satisfy equation \eqref{eq:appeom}; see \cite{gl16} for a discussion on the severity of this constraint. 

As before, the nonlinearity embedded within $L$ prevents a multipolar decomposition. We use the orthogonality relation \eqref{eq:orthog} to project the dipole equation for a function $f(r)$ introduced via $\psi(r,\theta) = f(r) \sin^2\theta$. This equation can be solved by imposing regularity at the center and matching to the external field \eqref{eq:bio6}, truncated to order $b^2$ here for simplicity. The Maxwell solution, which matches the analytic result from Appendix A of \cite{hask08} up to sign, is depicted in the left panel of Figure~\ref{fig:cst0}. Fixing $B_{p} = 2b/3$, the solution at order $b^2$ is shown in the right panel. Overall, the same qualitative features observed in Fig.~\ref{fig:mags} persist: (i) the field is locally stronger near the origin (by a factor $\sim 1.5$), and (ii) the region where $\psi > \psi_{c}$ effectively widens (where a toroidal field, if present, resides in the twisted-torus framework via expression \ref{eq:toroidalchoice}). 

\label{lastpage}

\end{document}